\documentclass{elsart}

\usepackage{graphicx}
 \usepackage{epsfig}
 \usepackage{subfigure}
\usepackage{amssymb}

\begin{document}

\begin{frontmatter}

% Title, authors and addresses

% use the thanksref command within \title, \author or \address for footnotes;
% use the corauthref command within \author for corresponding author footnotes;
% use the ead command for the email address,
% and the form \ead[url] for the home page:
% \title{Title\thanksref{label1}}
% \thanks[label1]{}
% \author{Name\corauthref{cor1}\thanksref{label2}}
% \ead{email address}
% \ead[url]{home page}
% \thanks[label2]{}
% \corauth[cor1]{}
% \address{Address\thanksref{label3}}
% \thanks[label3]{}

\title{Recent developments of surface light scattering as a tool for optical-rheology of polymer monolayers}

% use optional labels to link authors explicitly to addresses:
% \author[label1,label2]{}
% \address[label1]{}
% \address[label2]{}

\author[a]{Pietro Cicuta\corauthref{cor}}
\corauth[cor]{Corresponding author.} \ead{pc245@cam.ac.uk}
 and
\author[b]{Ian Hopkinson}
\address[a]{Cavendish Laboratory, University of Cambridge,\\
Madingley Road, Cambridge CB3 0HE, U.K.}
\address[b]{Department of Physics, U.M.I.S.T., Manchester M60 1QD, U.K}

\begin{abstract}
 Surface Quasi-Elastic Light Scattering (SQELS) is an application of dynamic light scattering  to measure the
dynamics of the thermal roughness of liquid surfaces. An analysis
of the spectrum of thermal fluctuations provides information on
 surface  properties like tension and elasticity. In this work we will focus
particularly on its use to study polymer or polymer-like Langmuir
monolayers. We review work in this area and give an up-to-date
overview of the method. Important advances have very recently
taken place in the theoretical understanding of this problem, and
this has allowed improvements in the analysis of the experimental
data. A practical method to estimate the region of physical
parameters that can be reliably measured is presented.
\end{abstract}

\end{frontmatter}
%\subsection{Motivation and history of the technique} \label{sqels motivation}
Many complex fluids, such as foams and emulsions,  are multi-phase
systems characterized by a very high interfacial area. Often their
response to deformation is determined by the properties of the
quasi-two-dimensional interfaces, and this makes  the study of
surface viscoelasticity  of great interest~\cite{buzza95}. Of
course sometimes the two-dimensional dynamics  is  itself the
object of interest. Relatively few experimental techniques exist
to probe surface rheology, and surface quasi-elastic light
scattering (SQELS)  is unique in many respects, such as being a
non-invasive and non-perturbative probe. Progress up to 1992 is
reviewed in Langevin's monograph~\cite{lan92}  and further work is
reviewed by Earnshaw in~\cite{earnshaw96}. These sources discuss
some of the points summarized in this manuscript in much greater
detail and contain references to original work. SQELS has been
used to study various {\it soft matter} systems, including
microemulsions, polymer solution interfaces
\cite{huang98,kubota98}, bilayer lipid membranes, soap films,
surfaces of liquid crystals or gelled systems.    In this paper we
review one particular use of the SQELS technique, its application
to investigating surfaces decorated by monolayers. The  dynamics
of monolayers of  fatty acids, biological lipids, synthetic
surfactants and polymers is itself a rich topic, of interest in a
wide range of problems that range from understanding biophysical
processes to engineering molecular self-assembly. This paper
explains how and under what conditions SQELS can be used as a
  micro-rheological probe of viscoelasticity in a  polymer (or polymer-like)
  monolayer. It will hopefully provide both a clear introduction and an up-to-date reference.

Light scattering from thermal fluctuations of a liquid surface,
capillary waves, was first predicted by Smoluchowski in 1908. It
was observed experimentally by light-scattering pioneers
Mandelstam (1913), Raman (1924) and Gans (1926) but only when
laser light sources became available  experiments could be
attempted to resolve the theoretically predicted \cite{levich62}
frequency spectrum. Katyl and Ingard first observed the spectral
broadening of light reflected from methanol and isopropanol
surfaces with a Fabry-Perot interferometer \cite{katyl67} and soon
after, by using heterodyne spectroscopy, they were able to resolve
the Brillouin doublet. They called the surface fluctuations {\it
thermal ripplons} and used an optical setup  very similar to that
still employed today. The combination of theoretical understanding
of surface fluctuation hydrodynamics with the heterodyne optical
setup provided a precious tool for the study of critical phenomena
close to phase transitions in `simple' fluids. The experiments by
Huang and Webb \cite{webb69} and Meunier \cite{meunier69},
determining the critical exponent for the surface tension as a
function of temperature close to the critical point,  were the
first in a long series of studies that used light scattering by
surface fluctuations to measure the interface tension. For these
investigators  this method was important because it enabled non
invasive measurements and allowed probing of very low tensions.
These two points have remained valid motivations for the use of
this technique, as it was further developed and extended to be
applied to progressively more complex fluids.

 The measurement of the spectral shape of  light scattered from a surface
 covered by a monolayer  often enables
 the determination of intrinsic rheological
parameters, in addition to the measurement of surface tension. As
this technique relies on observing the dynamics of thermal
fluctuations, it is, almost by definition, probing the system's
equilibrium state. The frequency of thermal surface roughness
fluctuations is typically around 10~kHz and   the amplitude  is of
the order of a few ${\rm \AA}$, hence although the frequency is
high the
 strain rate is  small. As in traditional bulk rheology often one wants
to probe the linear response regime. This is not guaranteed by
most other surface rheology techniques, from drop dilation to
surface shear
rheometers~\cite{miller96,dimeglio99,fuller99,zasadzinski02b}.
These techniques all rely on measuring the system response
following a macroscopic external perturbation, which can affect
the conformation and structure of a complex system.

 Section~\ref{sqels theory} of this paper describes the
 theoretical background which is necessary to
 analyze surface light scattering, that is
 extracting surface rheological parameters from the spectral shape of scattered
 light.
 Understanding the physical processes  that give rise to those parameters, for example
 explaining their dependence on concentration, is a separate problem that  will  depend
 on the particular system. This second level of analysis is outside the scope of this paper.
  The experimental setup is described
 in section~\ref{Cambridge  hardware setup}. The methods of SQELS data analysis are
presented in section~\ref{Fitting data}, where an original and
practical method is presented to estimate the range of
applicability of SQELS. Finally, in section~\ref{A model polymeric
monolayer}, data on monolayers of the synthetic polymer PVAc is
reported as an example of the application of these methods. This
system is chosen because it has been studied with SQELS by various
authors, and it represents a suitable standard for comparison with
the literature.

\section{Theory for data analysis on simple monolayers}\label{sqels
theory}
 The    theoretical derivation of the
spectrum of light scattered by thermal roughness  of a surface
decorated by a monolayer   is due to  Lucassen-Reynders and
Lucassen \cite{lucassen69} and Langevin and Bouchiat
\cite{langevin71}. Important considerations have been recently
published by Buzza \cite{buzza02} and have resulted in
  much more robust  data analysis. The case of a decorated surface
  is an extension of the theoretical  description of capillary waves on the clean surface of a
  fluid. As reviewed in \cite{lan92}, this topic was initially studied by Rayleigh and Mandelstam
  and was developed further by
   Levich  and  Papoular. The
  problem is to find a solution to the Navier-Stokes equations,
  which are linearized under the assumption of low velocity which
  holds  for waves with an amplitude small compared to the
  wave-length:
\begin{eqnarray}
\nabla\cdot {\bf v}\,=\,0,\\
  \rho \frac{\partial {\bf v}}{\partial
  t}\,=\,\nabla\cdot\sigma^{(1,2)},
 \label{navier1}
\end{eqnarray}
where the hydrodynamic stress tensor is
\begin{eqnarray}
\sigma^{(1,2)}\,=\,\eta_{(1,2)}\left(\frac{\partial v_i}{\partial
x_j }+\frac{\partial v_j}{\partial x_i } \right)\,-\,P\delta_{ij}
 \label{navier2}
\end{eqnarray}
and where {\it P} and ${\bf v}$ are the pressure and  velocity
fields. The flow has to satisfy the boundary conditions of
continuous velocity at the interface and of vanishing  {\it P} and
${\bf v}$ infinitely far from the interface. If the surface is not
flat, $P$ includes the Laplace pressure due to the surface
tension.

The same bulk hydrodynamic equations Eq.~\ref{navier1} and
Eq.~\ref{navier2} hold in the presence of a monolayer at the
interface. The surface film has usually been treated as a
mathematically thin plane characterized by a tension, elasticity
and viscosity, and it simply contributes additional terms to the
stress tensor equation~\ref{navier2}, as explained clearly in
\cite{lucassen69}.  Thick films or systems with complex
interactions with the subphase, like for example diffusion
exchange, fall outside of this approximation. Even for this
simplest case the process of finding the correct and general
constitutive model for interface motion has proven challenging,
and remained a long-standing controversial issue. Buzza has
shown~\cite{buzza02} that many models that have had over time a
profound influence  in the literature are incorrect. It is proved
rigorously that the parametrization by Scriven in terms of a
surface tension and complex dilational and shear moduli is the
most general possible, and  the model is extended to account for
relaxation processes in the monolayer. The same paper  also
clarifies how the models by Goodrich and Kramer \cite{kramer71}
erroneously led to consider a complex surface tension, the
imaginary part of which was called a transverse viscosity. As a
conclusion of Ref.~\cite{buzza02} a surface wave dispersion
relation $D(\omega)$ is found relating the wave frequency to the
wavelength. This expression is
 the same as the original based on Goodrich (see
\cite{langevin71}), but with the important difference that the
surface tension $\gamma$ is a real quantity and has to be equal to
the equilibrium static surface tension:
\begin{eqnarray}
D(\omega)\,&=& \nonumber \\  \,\left[\,\varepsilon^*
q^2\,+\,i\omega \eta \left( q\,+\,m\right)\,
\right]&\,&\left[\,\gamma q^2\,+\,i\omega\eta \left( q\,+\,m
\right)\,-\,\frac{\rho \omega^2}{q}\, \right]\,-\,\left[\,i\omega
\eta \left(m\,-\,q \right)\, \right]^2,\nonumber \\
 \label{dispeq}
\end{eqnarray}
where $m$ is
\begin{eqnarray}
m\,=\,\sqrt{\,q^2\,+\,i\frac{\omega \rho}{\eta}}, \,\,\,{\rm
Re}(m)>0,
 \label{eqform}
\end{eqnarray}
$\eta$ is the subphase Newtonian viscosity, $\rho$ is the subphase
density, $\gamma$ is the surface tension (or transverse modulus)
and $\varepsilon^*$ is the complex dilational modulus. Formally,
the in-plane modulus measured by SQELS is the sum of the elastic
dilation and shear moduli, however the shear modulus  is
negligible in many polymeric monolayers so the data will be
discussed in terms of the dilational modulus only.
 Solving
this equation for $D(\omega)\,=\,0$ gives  an expression for the
wave frequency, $\omega$, as a function of the wave-vector $q$.
The solutions describe both dilational (in-plane) and transverse
waves \cite{lucassen69}. In a surface light scattering experiment
it is only the transverse waves that scatter light, and  power
spectrum of scattered light is P$_q$($\omega$)  given by
\cite{langevin71}:
\begin{eqnarray}
P_q(\omega)\,=\,\frac{k_B T}{\pi \omega}{\rm Im}\left[\,\frac{i
\omega \eta (m\,+\,q)\,+\,\varepsilon^* q^2 }{D(\omega)}\,
\right].
 \label{powersp}
\end{eqnarray}
The behaviour of the dilational waves can be inferred from the
time evolution of transverse waves because the dynamics of the two
modes are coupled through Eq.~\ref{dispeq}. As will be shown in
section~\ref{Cambridge  hardware setup}, the heterodyne time
autocorrelation function which is measured by SQELS is the Fourier
transform of the power spectrum given by Eq.~\ref{powersp}.

The important practical consequence of Buzza's analysis is that if
the surface tension $\gamma$ is measured independently (and
simultaneously to ensure the necessary precision), then only two
physical parameters $\varepsilon$ and $\varepsilon'$ need to be
fitted from the experimental data. This is very different from the
data analysis of early SQELS experiments, where four fitting
parameters where usually considered, and provides a general
justification for the correct recent practice (which was motivated
by the theory proposed for a specific case \cite{buzza98}) of
considering a purely real surface tension. In summary, fitting
with a free surface tension parameter should be attempted only if
the surface tension cannot be reliably measured independently.

\section{Experimental setup}\label{Cambridge  hardware setup}
This section describes  the experimental  aspects of SQELS and the
experimental setup used in our laboratory, which  is based on the
instruments described by Earnshaw~\cite{earnshaw87} and H{\aa}rd
and Neuman~\cite{neuman81}.
%
%
%\\ FIGURE -diagram of sqels\\
\begin{figure}[t]
%\begin{center}
\epsfig{file=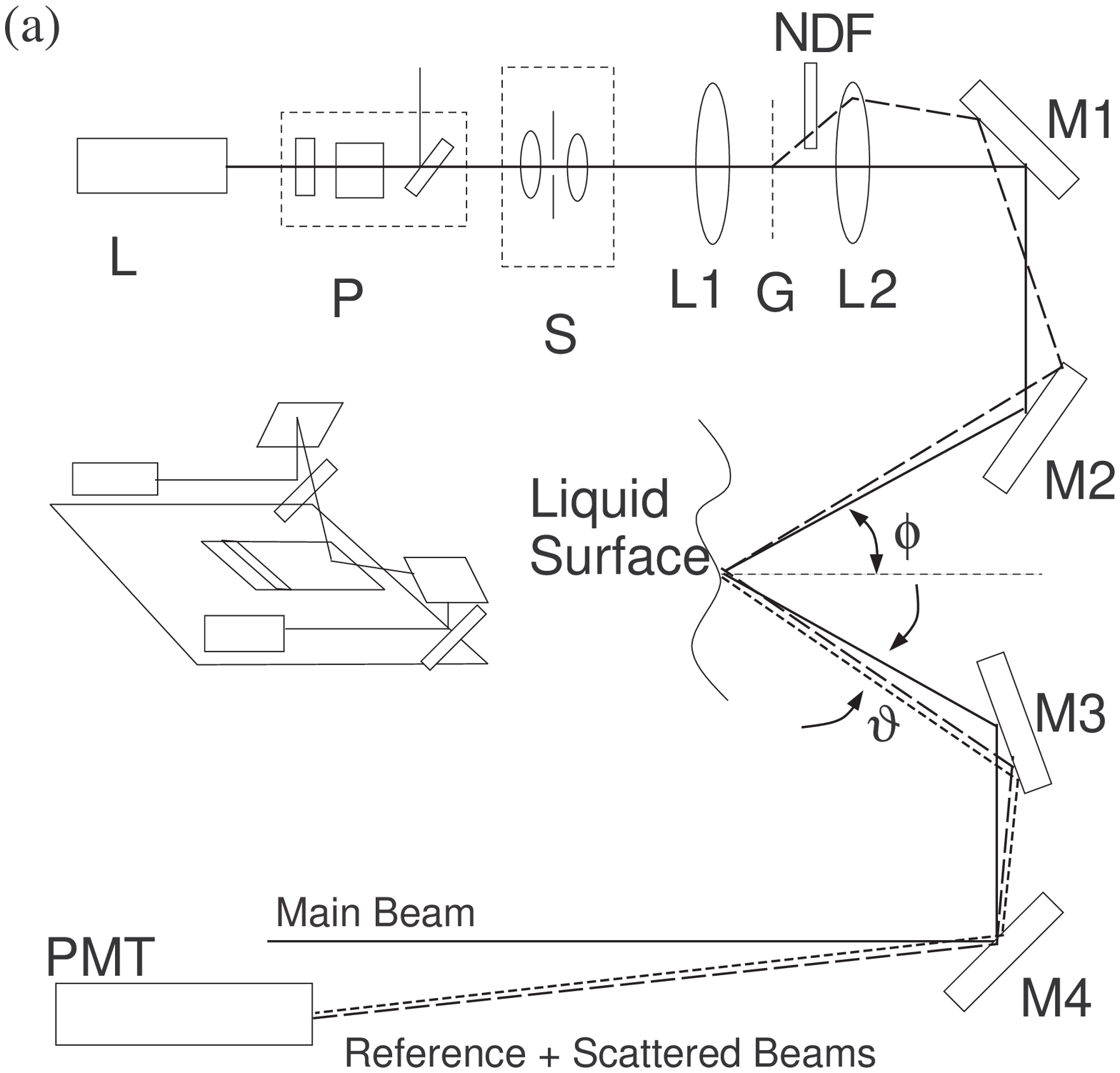, width=6.5cm}
\epsfig{file=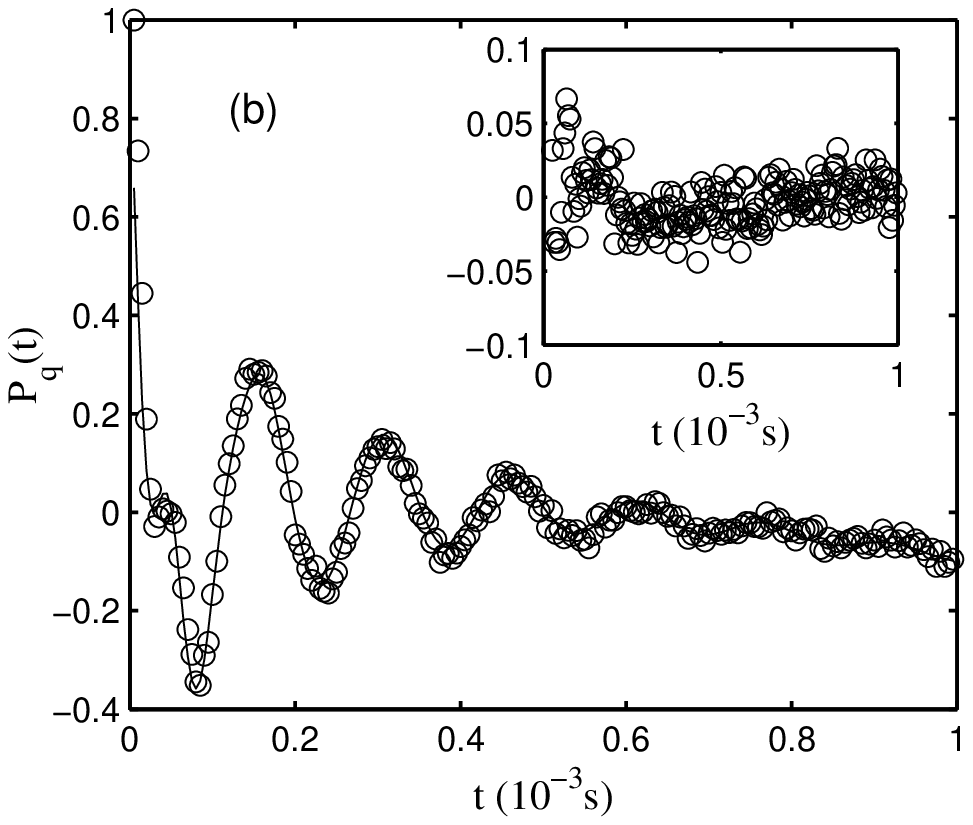, width=7cm} \caption{(a)~Diagram of
the SQELS optical setup. Each part of the optical train is
described in the text. The diagram clearly shows the main beam
(solid  line), one of the reference beams created by the grating G
(dashed  line) and the scattered field (dotted line). (b)~A
correlation function obtained by SQELS at $q=306$cm$^{-1}$ on a
surface decorated by a PVAc monolayer at surface pressure
$\Pi=3.7$mN/m, water subphase, T=25$^\circ$C. The solid line is a
fit with equation~\ref{powersptime} as described in the text, and
the inset shows the residuals from this fit. \label{sqels setup
figure}}
%\end{center}
\end{figure}

\subsection{Optical train}
 Figure~\ref{sqels
setup figure}  illustrates the instrument schematically. Two
different light sources (L) have been used: a  150mW single mode
diode pumped solid state laser (Laser Quantum, U.K.) with a
wavelength $\Lambda=532$nm and a 30mW He-Ne gas laser (Newport,
U.S.A.) with a wavelength $\Lambda=633$nm.
      For both lasers, the polarization (horizontal) and intensity are controlled
using  the combination of a half wave plate  and prism polarizer
(P). A  combination of prism polarizer and neutral density filters
can also be used instead. The beam size, profile and collimation
are controlled using a spatial filter, S, composed of two
microscope objectives and a pinhole. The position of the expansion
lens is such that a Gaussian (parallel) beam emerges from the
filter assembly, and the choice of focal length determines the
beam waist. An appropriate choice of beam diameter has to be made,
depending on the physical parameters of the liquid surface under
investigation. The optimal condition is detection of scattered
light from an area of the surface of the order of the capillary
wave coherence length squared. Instrument resolution will become
very bad  if the illuminated area is much smaller, whereas the
signal to noise ratio will fall if the area is too big, in which
case one is averaging over uncorrelated dynamics \cite{pecora76}.

Photon correlation is done in heterodyne mode (as discussed in
detail below), so it is necessary to provide a coherent source of
light of the original frequency at the appropriate scattering
angle. This is done with a weak diffraction grating (G) which
 provides a fan of diffracted beams  serving as
`reference'. The relative intensity of the reference beams is
adjusted by inserting a neutral density filter (NDF)  that
intercepts the diffracted spots but not the main beam. The lenses
L1 (f=150mm) and L2 (f=350mm) perform two tasks; they converge the
fan of reference beams and the main beam to a single spot at the
fluid interface and they focus both the reference beams and the
main beam in the front plane of the photomultiplier, situated
about~2m after the surface. In order for the heterodyne signal to
dominate the correlation function, the ratio of the intensity of
the inelastically scattered light to the `reference' light must be
adjusted to a value of the order of 10$^{-3}$ \cite{earnshaw97}.

 The mirrors
M1-M2 direct light from the laser onto the liquid surface. A
fraction of the light is refracted into the liquid and, to reduce
stray light, a mirror on the trough bottom deflects it away from
the forward direction. Most of the light is reflected specularly
from the surface and a small fraction is scattered by the thermal
roughness. The mirrors M3-M4 collect the cone of light in a small
solid angle around the main beam, and the mirror M4 is adjusted so
that the reference beam (together with the scattered field) is
directed into a 1mm diameter aperture in front of the
photomultiplier tube.  The optical train described above is
mounted onto a standard optical table. The liquid surface is
contained in a trough resting on an active-damping anti-vibration
platform~(Mod-2S, Halcyonics, Germany) and situated in a
draft-proof enclosure. With the isolation unit active, the laser
beam reflected from the liquid surface appears very stable.

\subsection{Detected signal} For light polarized in the plane of incidence,
the scattering angle $\theta$ is related to the capillary wave
vector $q$ by:
\begin{eqnarray}
q\,=\,\frac{2 \pi}{\Lambda}\cos(\phi)\sin(\theta),
 \label{sqelshard1}
\end{eqnarray}
where $\phi$ is the angle of incidence on the liquid surface  and
$\Lambda$ is the wave-length of the laser light.
   The time averaged scattered intensity per solid angle is
given by \cite{lan92}:
\begin{eqnarray}
\frac{dI}{d\Omega}\,=\,\frac{I_r(\phi)}{A}\frac{16
\pi^2}{\Lambda^4}\,\cos^3(\phi)\,\langle | h_{\bf q} |^2\rangle_t,
 \label{sqelshard2}
\end{eqnarray}
where $I_r(\phi)$ is the reflection coefficient and $\langle |
h_{q} |^2\rangle_t$ is the mean square amplitude of the mode $q$
surface roughness, which can be easily calculated for an interface
in thermal equilibrium~\cite{lan92}:
\begin{eqnarray}
\langle | h_{\bf q} |^2\rangle_t\,=\,\frac{1}{A}\frac{k_B
T}{\gamma q^2}.
 \label{roughness}
\end{eqnarray}

 Light is detected using a
photomultiplier tube (PMT). At the detector the specularly
reflected laser beam appears as a central spot, surrounded by a
halo of inelastically scattered light (from the liquid surface)
and a horizontal series of focused reference spots at 2-3mm
intervals away from the central spot.
%
%
%\begin{figure}[tbp]
%         \begin{center}
%         \epsfig{file=./figch2/correlation.ps,width=8.5cm}
%       \end{center}
% \caption{Typical correlation functions obtained by
%surface quasi--elastic light scattering  at subphase
%T=22$^\circ$C, $p$H=5.24 and scattering vector $q$=425cm$^{-1}$
%under different conditions (a): `bare' buffer; (b):
%$\beta$--casein monolayer at concentration $\Gamma=1\times
%10^{-3}$g/m$^2$ and pressure $\Pi=6.3\times 10^{-3}$N/m. The solid
%lines are fits to the model given in Eq.~\ref{powersptime}, and
%insets show the residuals of the fits. \label{correlation}}
%\end{figure}
%
%\subsubsection{heterodyne light beating}
 The
total intensity of light falling onto the PMT is composed of
 the reference $E_r(t)$ and  the scattered $E_s(t)$ light
fields \cite{earnshaw97}:
\begin{eqnarray}
I(t)\,=\,I_s(t)+I_r(t)+2{\rm Re}[E_s(t)\,E_r^*(t)].
 \label{sqelshard3}
\end{eqnarray}
 The signal $I(t)$ is  processed using a PC-card based
photon correlator (BI9000, Brookhaven Instruments, USA). A pulse
discriminator is used in the PMT and has been factory-modified to
allow the use of the `multi-photon' mode, originally described by
Winch and Earnshaw \cite{earnshaw88}, to increase the sensitivity.
The correlator measures the autocorrelation of $I(t)$:
\begin{eqnarray}
P(\tau)\,=\,\langle I(t)\,I(t+\tau) \rangle_t\,=\,(I_s+I_r)^2+2I_s
I_r g_s^{(1)}(\tau)+I_s^2[g_s^{(2)}(\tau)-1],
 \label{sqelshard4}
\end{eqnarray}
where
\begin{eqnarray}
g_s^{(1)}(\tau)=\frac{\langle E_s(t)\,E_s(t+\tau)
\rangle_t}{\langle I_s(t) \rangle_t}-1 \,\,
   {\rm and}\,\,
g_s^{(2)}(\tau)=\frac{\langle I_s(t)\,I_s(t+\tau)
\rangle_t}{\langle I_s^2(t) \rangle_t}-1
 \label{sqelshard6}
\end{eqnarray}
are respectively the first and second order normalized
autocorrelation functions of the scattered field.
$g_s^{(1)}(\tau)$, which is the same as the Fourier transform of
the  optical spectrum, is the object that needs to be measured
\cite{pecora76}. This is possible for $I_r \gg I_s$, when the
second term of Eq.~\ref{sqelshard4} dominates the time dependance
in the autocorrelation function. Figure~\ref{sqels setup
figure}(b) shows a typical correlation function with the baseline
subtracted and with the amplitude normalized to~1.

\subsection{Practical notes}
 The reference beams are
sufficiently weak that the inelastic scattering from them can be
ignored, and each reference spot corresponds to light being
scattered at a different $q$ value. They were first introduced in
this geometry by \cite{hard76} and serve principally to
 allow control over the relative intensity of the light fields
(thus assuring good heterodyne conditions), to
    enable repeated
experiments at very reproducible scattering angles and to
    reduce the sensitivity of the setup to surface sloshing which can
result in slow oscillations of the reflected reference beams and
the scattered field.

    In the common use of the instrument, the modulus of the
scattering vector, $q$, and the instrument resolution at a
particular reference spot are calibrated at the beginning of each
experimental session by fitting the measured correlation function
for a pure liquid (typically water) at that scattering angle.
During an experiment, and depending on the system studied,
various correlation functions are acquired consecutively under the
same conditions, with each correlation function accumulated over
typically a few minutes. SQELS data can be acquired from
monolayers maintained in a Langmuir trough or in other containers
such as Petri dishes, provided the diameter is big enough that a
central region of the surface is sufficiently flat. For a
water/air interface, below a diameter of 6cm the de-focusing
effect of the curved interface on the reflected beam is very
noticeable.

\subsection{Instrumental broadening}\label{Instrumental
broadening} Soon after the first observation of the  broadened
power spectrum of light scattered by capillary waves  it was
recognized that the observed frequency width was inconsistent with
the predicted signal for clean liquid surfaces. It took until
\cite{langevin74,hard76} to develop a way of separating out the
instrumental broadening due to the optical resolution from the
physical signal. The discussion continued in
\cite{shih84,mann84,swofford88} on how to  take into account the
beam propagation through the optical train and on how to optimize
the   optical design  to maximize the resolution. These issues are
truly essential only if the instrument is to be used to give
absolute values of the (unknown) surface parameters of a certain
liquid (of which the bulk parameters have to be known anyway). In
the use of the technique to study monolayers, an easier working
practice is
 appropriate: The
instrumental parameters (scattering angle and resolution) are
calibrated by acquiring the spectrum of scattered light from the
clean interface of the subphase, for which both bulk and surface
parameters are usually known precisely. Subsequent measurements
are performed after spreading the monolayer without any changes
taking place on the instrument. The instrumental broadening
function has been shown to be adequately approximated by a
gaussian form as discussed in~\cite{earnshaw90b}. This allows the
resolution to be described by a single parameter, $\beta$. In very
accurate measurements an improvement could be detected by
describing broadening with a two-parameter Voight
function~\cite{langevin74}. We find the gaussian broadening
correctly accounts for the resolution in our measurements.

\begin{figure}[t]
 \epsfig{file=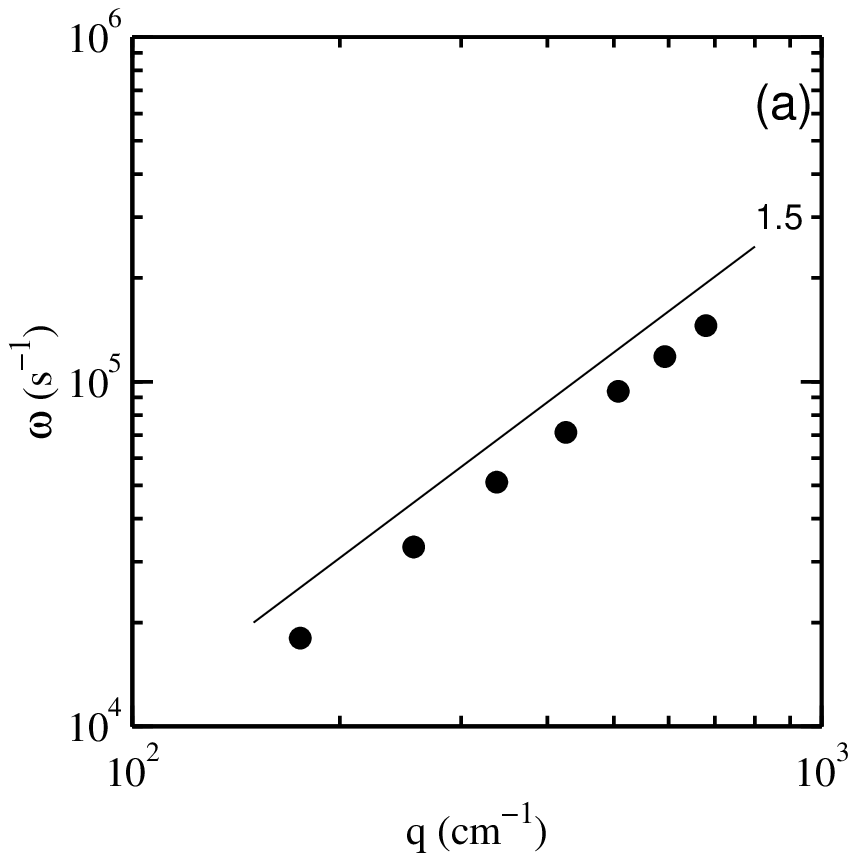,height=4.5cm}
 \epsfig{file=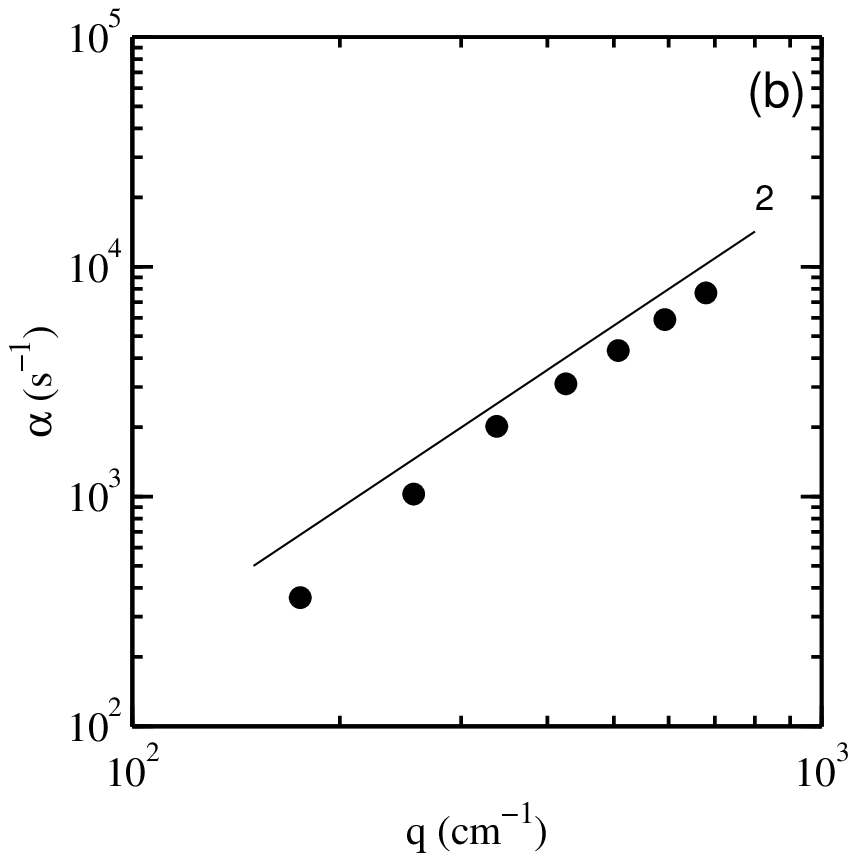,height=4.5cm}
  \epsfig{file=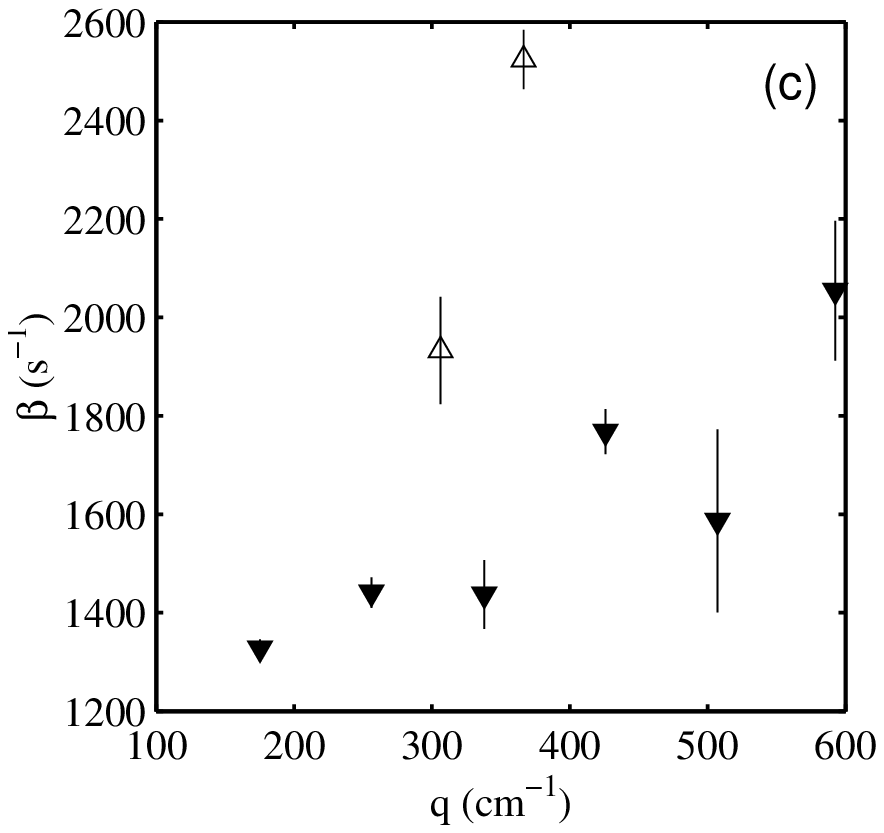,height=4.5cm}
 \caption{ Log-Log plots of (a) the frequency and (b) the damping factor of capillary waves as a function of
     the wave-vector, obtained by fitting data with Eq.~\ref{dampedcosine}.
      Data is obtained on a clean water surface at 22$^\circ$C.  The solid lines are
      approximated solutions to Eq.~\ref{dispeq}, known as the Kelvin and Stokes laws~\cite{lan92}.
(c)~Instrumental resolution (parameter $\beta$ in
Eq.~\ref{powersptime}) as a function of the
 scattering wave-vector. ($\blacktriangledown$):~data acquired with the 532nm laser beam, and
 ($\vartriangle$):~data are typical of the setup with the 633nm laser. The resolution
 depends strongly on the laser beam waist. \label{sqelssystemres}}
\end{figure}

\subsection{Other hardware configurations} The SQELS setup
described above has evolved over many years. Numerous groups have
tested and compared different geometries and hardware options, and
at the same time  purely technological improvements in electronics
(the digital correlator and active anti-vibration isolation) and
lasers have taken place. It is particularly suited to studying
monolayers at the air-liquid surface.

%\subsubsection{Modifications}

 Alternative approaches include a vertical incidence reflection geometry, with a combination of
polarized beam-splitter and  quarter-wave plate to separate
incident and reflected light beams as presented in
\cite{swofford88}.  A transmission geometry was employed by
\cite{takagi91}. This setup is also original in the use of an
acousto-optic modulator to provide the reference signal, which
enabled access to scattering vectors up to
$3\times10^{4}$cm$^{-1}$, one order of magnitude higher than the
maximum attainable with traditional systems \cite{earnshaw93}.
Other setups that make use of  a transmission geometry are
\cite{jorgensen92,sawada97,meyer97,meyer97b,trojanek01,meyer01a}.
A transmission geometry has the advantages of requiring a simpler
optical train and  reduces the  sensitivity of the technique to
slow oscillations ({\it sloshing}) of the surface. Other
approaches to minimizing sloshing are discussed
by~\cite{meyer01b}.  The main limit of the transmission geometry
is that only systems with a negligible scattering from the liquid
subphase can be studied.

The specific  issue of   cost in building a surface scattering
setup was addressed in \cite{jorgensen92}, where the possibility
of using a cheap light source (multi-mode diode laser) and
intensity detection (silicon photodiode) is explored.  A further
simplification  is to use optical fibers instead of the optical
train, as in the setup described in \cite{meyer97}. This has the
additional advantage of enabling a more compact and robust
instrument. Thanks to progress in computer power, in the near
future it should become possible to substitute the digital
correlator with a (relatively inexpensive) dedicated data
acquisition board, feeding data to a personal computer and
performing real time software based correlations \cite{ferri01}.
As a testament to the maturity of the technique, it is significant
to reference an educational paper~\cite{yu02}  where SQELS is
suggested as an appropriate experiment in an undergraduate student
course.

%\subsection{Presently active  research groups}
Various research groups are actively studying polymer monolayers
with SQELS. We review this research over the past ten years, after
the publication of Langevin's monograph~\cite{lan92}. The group of
Richards  has done extensive work on polymer monolayers,
 some of which is covered in the monograph~\cite{richards99}. Recent measurements on multiblock
 copolymers are reported in \cite{richards00}.
The group of Monroy, Ortega and Rubio  has worked on monolayers of
polymers and polymer mixtures. In their work monolayer parameters
are extracted
 from the  fitted values of frequency and
damping obtained from the power spectrum, an  approach which is
critically discussed  in section~\ref{Fitting data}. SQELS has
been used to study the viscoelasticity of   model polymer
monolayers such as  PVAc \cite{monroy98,monroy00b}, PVAc and P4HS
blends \cite{monroy99}, and P4HS \cite{monroy01}. In addition to
capillary wave data, this group has, for some systems, presented
comparative measurements obtained by different techniques
\cite{monroy98,monroy01},  providing valuable data spanning a very
wide frequency range. The  group of Yu also studied model polymers
representative of good and $\theta$ solvent
conditions~\cite{yu89}. Diblock copolymers are studied in
\cite{yu94}. In a  recent paper~\cite{yu00} light scattering
results on monolayers of six different polymers are reviewed and
compared.

%There are recent examples of SQELS being used on monolayers simply
%as a technique to measure the surface tension from the frequency
%of scattered light. The  group of Sawada in Japan has published a
%series of papers, most recently \cite{sawada97}, in which the
% time-evolution of the surface tension is determined. A similar use
%of the technique by a different group is \cite{trojanek01}.

\subsection{Technical details}
Figure~\ref{sqelssystemres} shows the range of scattering vectors
$q$ that can be measured with the SQELS apparatus. The values of
the corresponding capillary wave frequency
(Figure~\ref{sqelssystemres}(a)) and damping constant
(Figure~\ref{sqelssystemres}(b)) are shown in log plots. The
slopes, as expected in this range~\cite{earnshaw97}, are in
agreement with the first order approximations that can be made to
Eq.~\ref{dispeq}, as is well documented in the literature
\cite{lan92}.
%
%\begin{figure}[tpb]
%   \centering
%        \subfigure[]{            \label{sqelssystem:subfig1}
%        \epsfig{file=./FIGCH2/systemtamA.ps,height=7cm}
%         }
%          \subfigure[]{            \label{sqelssystem:subfig2}
%        \epsfig{file=./FIGCH2/systemtamB.ps,height=7cm}
%         }%% \qquad
%     \caption{ Log-Log plots of (a) the frequency and (b) the damping factor of capillary waves as a function of
%     the wave-vector.  Red, green and blue points correspond respectively to a clean water surface and to
%     a $\beta$-casein monolayer at surface pressure $\Pi$=2mN/m and 6mN/m. The solid lines are
%      approximated solutions to Eq.~\ref{dispeq}, known as the Kelvin and Stokes laws~\cite{lan92}.\label{sqelssystem}}
%\end{figure}
%
The instrumental resolution values that are measured for different
scattering angles are shown in Figure~\ref{sqelssystemres}(c),
where it is clear that the instrument setup employing the 533nm
laser had a considerably better resolution than the 633nm laser.
The setups also had slight differences in positioning the optical
train, so that  different beam waists were obtained.

It is  well known  that  a  range of time-delay channels on the
digital correlator  should be selected so that the full extent of
the decay of the correlation function is measured. However it can
happen in SQELS experiments that for some scattering angles it is
difficult to find an optimum match of channel times to correlation
decay. With our setup, this happens for the smallest scattering
angles. It was found that if the channels were chosen as to
cut-off the correlation function, this had a {\it very} strong
effect when fitting the data with our software program, described
below. This is because the discrete Fourier transform routine
assumes periodic boundary conditions on the data (other routines
pad the data with zeros, which would also lead to the same
problem). Hence, if the data has not decayed to a flat baseline,
there is the possibility that the correlation function, which is
approximately a damped cosine oscillation, does not match the
periodicity of its symmetric image at large times. This problem
has a characteristic signature: a bad fit and  large residues are
found at the largest correlator times. When such flawed data is
fitted,  it leads to either negative values of the resolution
parameter or to negative values of $\varepsilon'$. In numerous
papers in the literature it is clear that the time correlation
data has been cut off, and this problem is not acknowledged.
Usually, the cut-off does not hide important features of the
correlation function, and the data can be used. However whilst
fitting, it is essential to ensure that the correlation function
connects with its symmetric image without loss of periodicity, for
example by careful choice of the number of channels that are
analyzed.

\section{Data Analysis}\label{Fitting data}
\subsection{Direct spectral analysis}\label{Direct spectral analysis} The maximum information
that can be extracted on the surface parameters of a monolayer
from the time correlation (or the frequency power spectrum) of
light scattered from capillary waves is obtained by fitting the
data directly, with the theoretical model given in
Eq.~\ref{dispeq} and~\ref{powersp}. As discussed in
section~\ref{Cambridge hardware setup}, our setup measures the
time correlation function, which is related to Eq.~\ref{powersp}
by~\cite{earnshaw90b}:
\begin{eqnarray}
P_q(t)\,&=&\nonumber \\
\,B\,&+&\,A\,{\rm F.T.}\left[ P_{q}(\omega)\right]\, \exp(-\beta^2
t^2/4)\,+\,{\rm droop}\,+\,{\rm fast\, afterpulse},
 \label{powersptime}
\end{eqnarray}
with $A$ and $B$ instrumental parameters and where the  `droop'
and `after-pulse' terms are added to correct for slow surface
oscillations and fast phototube after-pulse respectively. They
have the forms:
\begin{eqnarray}
 {\rm droop}\,&=&\,-d\,t^2,  \label{droop and after} \\ \nonumber
 {\rm fast\, afterpulse}\,&=&\,f_1\cos[f_2\,t]\,\exp[-t/f_3].
\end{eqnarray}
%  y(j)=+x(6)*exp(-x(7)*x(7)*t(j)*t(j)/4.0)*(cr(j))+x(8)*t(j)*t(j)+x(5) !Spectrum*resolution*amplitude+ droop+background
%   y(j)=y(j)+x(9)*cos(x(10)*t(j))*exp(-t(j)/x(11)) !+Fast process

We follow Earnshaw's approach to data analysis~\cite{earnshaw90b}.
Our fitting program is coded in \textsc{Fortran}. In essence, a
non-linear curve fitting of the model Eq.~\ref{powersptime} to the
 SQELS time correlation data  is carried
out in the time  domain, where analytical approximations to
account for instrumental parameters such as broadening, `droop'
due to surface sloshing and the amplitude of a fast phototube
`after-pulse' are known (Eq.~\ref{droop and after}). However, the
theoretical model Eq.~\ref{powersp} gives the power spectrum of
light in the frequency domain, so at every fitting step the
Fourier transform of the model data is performed, and then  the
instrumental corrections are applied. Rather than describing in
greater detail the fitting program, which  is very similar to
Earnshaw's, it is worth presenting  our fitting methodology, and
how it has improved since the original presentation.

\begin{figure}[t]
    \centering
        \epsfig{file=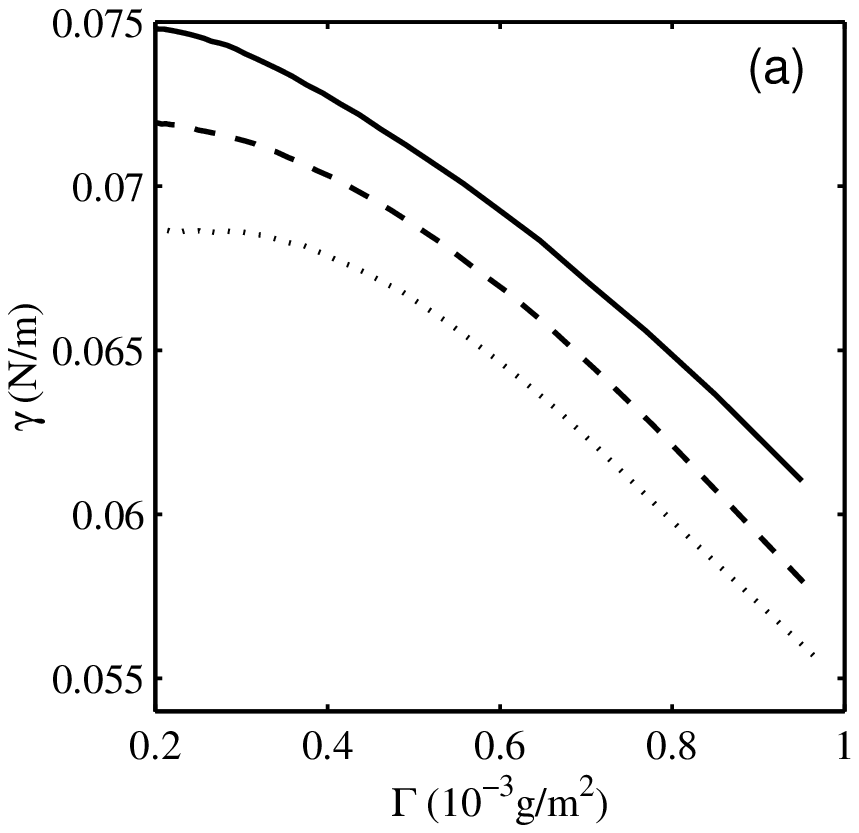,height=4.5cm}
         \epsfig{file=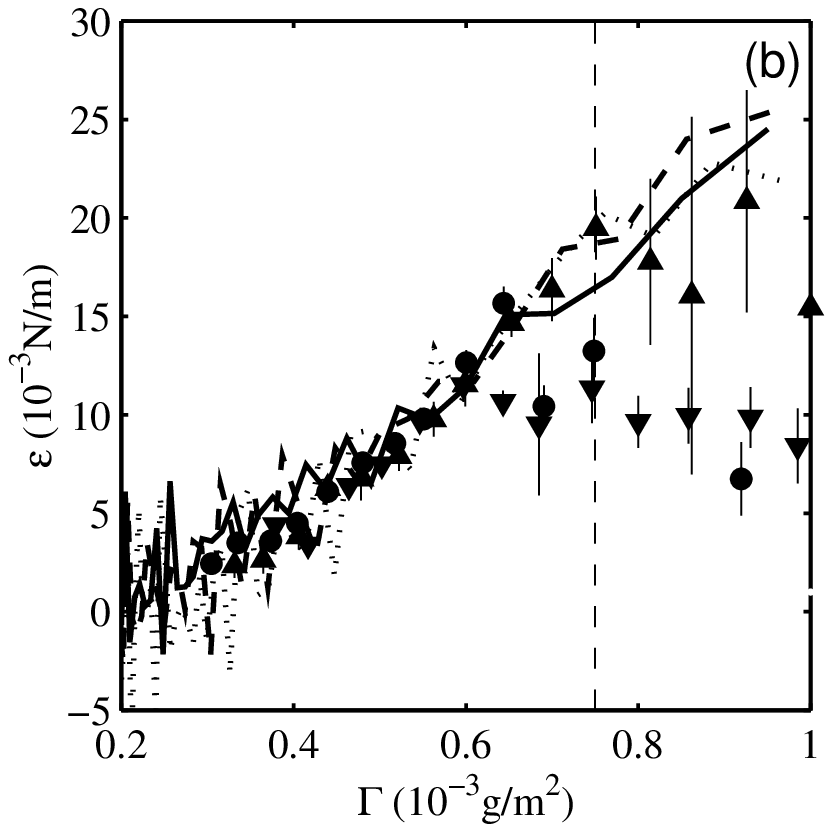,height=4.5cm}
          \epsfig{file=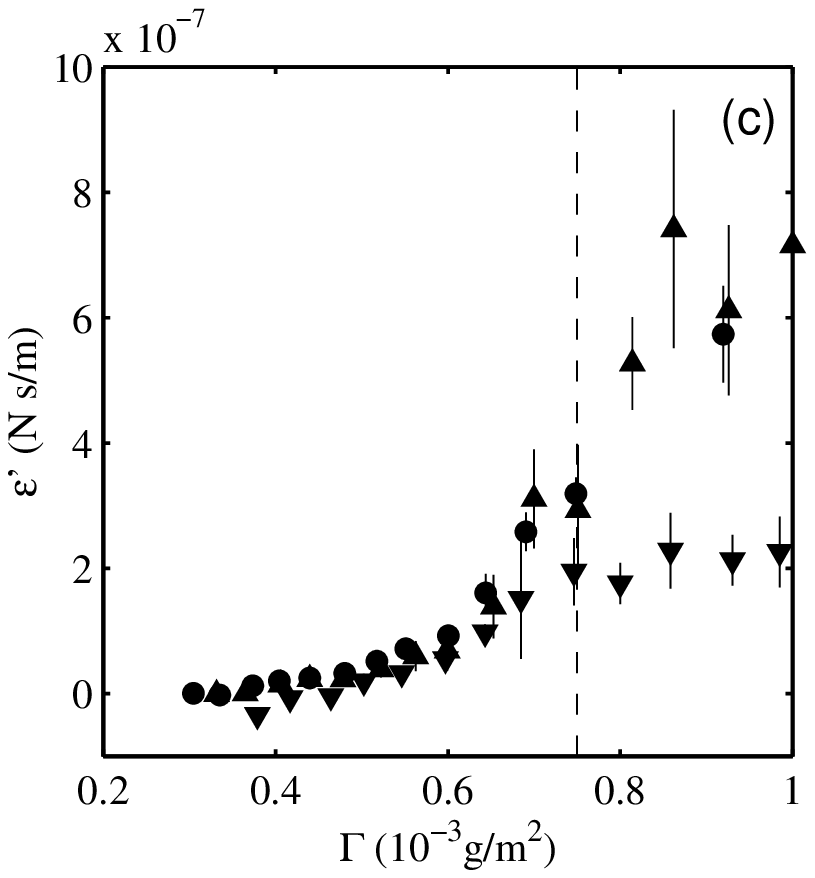,height=4.8cm}
       \caption{Surface parameters of  PVAc monolayers. (a)~Surface tension,
       measured with a Wilhelmy plate. Solid, dashed and dotted lines
       correspond respectively to 6$^\circ$C, 25$^\circ$C and 45$^\circ$C. (b)~Dilational
       elastic modulus and (c)~dilational viscosity, obtained by fitting SQELS data  with  Eq.~\ref{powersptime}. Symbols
       correspond to the subphase temperature:~($\blacktriangle$):~T=6$^\circ$C,
       ($\bullet$):~T=25$^\circ$C and ($\blacktriangledown$):~T=45$^\circ$C.
       The lines in (b) are the
equilibrium dilational moduli obtained
 from  isotherm experiments. As discussed in the
text, the surface tension is fixed to the static value during
SQELS data analysis. Initial fit parameters are
$\varepsilon=10^{-3}$N/m and $\varepsilon'=10^{-8}$Ns/m. In the
low concentration region, no systematic effect was noted by
fitting with other reasonable initial values. In the high
concentration region,
 values of $\varepsilon$ less than equilibrium, which is un-physical~\cite{buzza02}, resulted even with
very high initial values. The vertical dashed lines delimit the
regions where the fitted values become un-physical,  see the text
and Figure~\ref{fitting region}.    \label{figuresqelsbcas2}}
\end{figure}

The most important point  is that it is now clear that some of the
parameters introduced in \cite{earnshaw90b} can be measured
independently and should thus not be subject to fitting when
analyzing an unknown monolayer. In particular the resolution
parameter $\beta$ in Eq.~\ref{powersptime} (which measures the
instrumental broadening discussed in section~\ref{Instrumental
broadening}) is strongly correlated to the parameters determining
the spectral shape, despite having a different functional form.
This  leads to very unstable fitting unless $\beta$ is separately
determined as a fixed instrumental parameter. It can be estimated
by analyzing data from a clean water surface, together with the
best value for the scattering vector $q$.

Given Buzza's conclusion that the surface tension is to be fixed
to the static value  (and if the tension is accessible to
measurements by other means), then  only the following parameters
are free to be determined from the fit of an experimental
correlation function with Eq.~\ref{powersptime}:
\begin{itemize}
    \item Two ``physical parameters'':
         $\varepsilon$
        and $\varepsilon'$
    \item Four ``signal corrections'': The amplitude {\it A}, background {\it B},
      droop coefficient {\it d} and
       amplitude of fast phototube after-pulse {\it
        f$_1$}. The parameters {\it f$_2$} and {\it f$_3$} are determined separately by
        fitting the phototube signal of light directly from the
        laser.
\end{itemize}
We measure the surface tension with the Wilhelmy plate method,
simultaneously to the acquisition of the light scattering signal.
Values of $\varepsilon$ and $\varepsilon'$ obtained by fitting
data directly are shown in Figure~\ref{figuresqelsbcas2}(b)
and~(c). We will discuss the range over which such fits are valid
in section~\ref{Estimating the valid fitting region}.

\subsection{Phenomenological fitting}\label{Phenomenological fitting}
  An alternative approach to data analysis is to fit the time correlation
function with a damped cosine function:
\begin{eqnarray}
P_q(t)\,=\,B\,+\,A \cos(\omega t \,+\,\phi) \exp(-\alpha
t)\exp(-\beta^2 t^2/4),
 \label{dampedcosine}
\end{eqnarray}
plus the the droop and after-pulse terms of Eq.~\ref{droop and
after}. This expression, having three parameters in addition to
the four signal corrections described above, provides an excellent
approximation of the correlation function calculated using the
dispersion relation \cite{richards99,buzza02}. As with the data
analysis described in section~\ref{Direct spectral analysis}, the
final term in
 Eq.~\ref{dampedcosine} is the instrumental
broadening, which should be calibrated separately. $\omega$ and
$\alpha$ are the frequency and damping constant of the capillary
wave oscillations, and $\phi$ is a phase term that accounts for
the deviation of the power spectrum from a Lorentzian form. No
knowledge of the nature of the interface is required for this
analysis, and the fitting procedure is very stable. This is in
fact the only meaningful analysis of the ripplon spectrum that is
possible if one does not have a model that provides a dispersion
equation to relate the ripplon spectrum to the microscopic
 surface moduli \cite{buzza98}, a situation that occurs for example for a
 heterogeneous monolayer~\cite{cicuta02}.
  In the literature, it is often
 found that
  data is fitted with  Eq.~\ref{dampedcosine}, and
 then either the fitted frequency and
damping  are  quoted (seldom are values of the phase $\phi$
reported), or a search for interfacial properties which would be
consistent with these values is attempted. Ref.~\cite{yu00}
 provides a very good  explanation of this method. However this approach
is not ideal. Firstly, all three parameters $(\omega,\alpha,\phi)$
should in general be reported, as they are needed to correctly
interpolate the data. Secondly if,  using the results of this fit,
the search for values of the surface parameters that are
consistent through Eq.~\ref{dispeq} with the fitted
$(\omega,\alpha,\phi)$ is attempted, then this is an unnecessary
approximation to testing Eq.~\ref{dispeq} directly on the data
with the direct spectral analysis fit described in
section~\ref{Direct spectral analysis}.

\begin{figure}[t]
         \epsfig{file=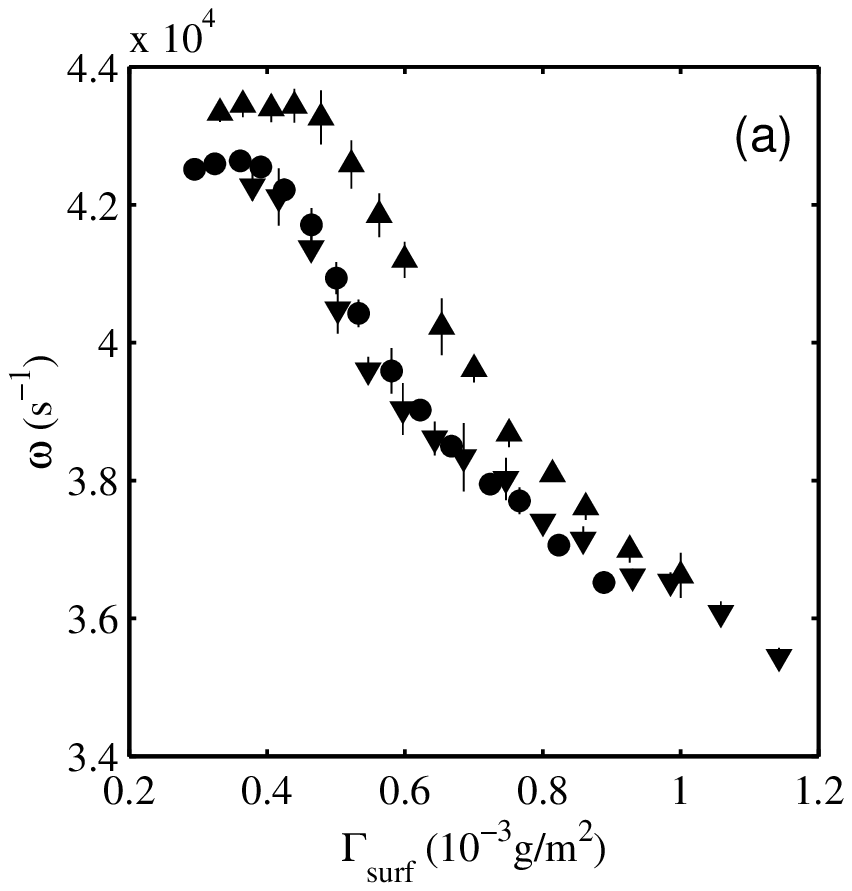,height=4.7cm}
         \epsfig{file=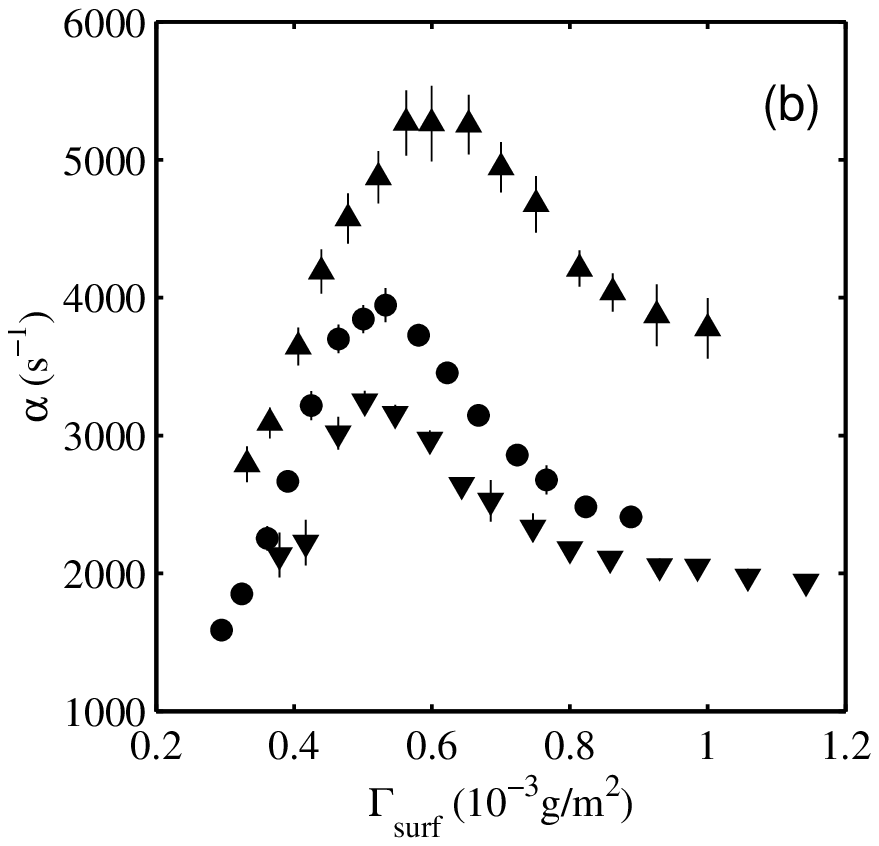,height=4.5cm}
         \epsfig{file=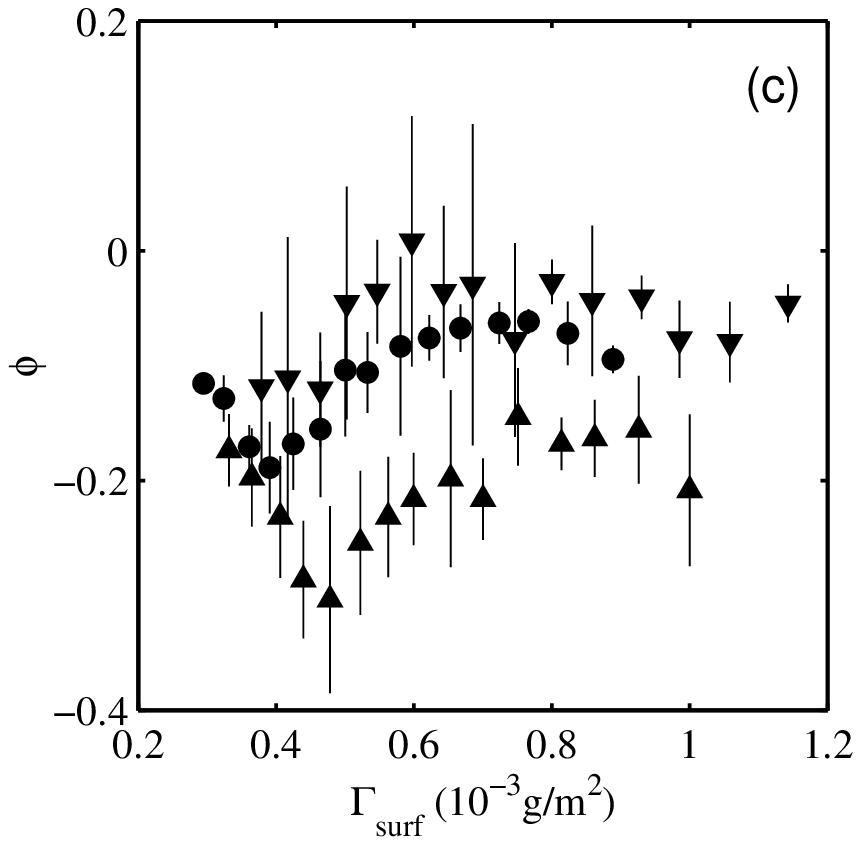,height=4.5cm}
    \caption{Phenomenological parameters of  PVAc monolayers at different temperatures, determined
    by fitting SQELS correlation functions with
    Eq.~\ref{dampedcosine}:
   (a)~Frequency, (b)~Damping
Coefficient, (c)~Phase term. The symbols correspond to the same
conditions described in
Figure~\ref{figuresqelsbcas2}.\label{figuresqelstam}}
\end{figure}

 In conclusion, the most reliable
surface parameters should result from fitting the data directly,
with Eq.~\ref{powersptime}, when this fit is stable. In the next
section it will be shown that the fit will work well only in a
limited region of physical parameters. Outside of this region, the
fitted values of Eq.~\ref{dampedcosine} become the only data that
can be discussed. Although a good fit with Eq.~\ref{dampedcosine}
is always obtained, one should not be misled: This approach too
will yield information on $\varepsilon$ and $\varepsilon'$ only in
the same limited range that will be estimated below. Indeed for
large $\varepsilon$ and $\varepsilon'$ the phenomenological fitted
values  saturate to a constant limit~\cite{yu00} which only
depends on the surface tension. The only advantage of the
phenomenological fitting approach is that the fit remains stable
over the whole concentration range, whereas the fit with the
physical parameters (Eq.~\ref{powersptime}) becomes unstable (and
can yield un-physical values of the dilational parameters) as the
boundary of the region of strong coupling between in-plane and
out-of-plane modes is approached.

%
%\\ TABLE - errors\\
%
%
\begin{table}[tpb]
\begin{center}
\begin{tabular}{|c||c|c|}  \hline
    Parameter  &   \,       Average conditions\,          &        \,Best conditions\,   \\
    \hline \hline
     $q$\,&\,             306cm$^{-1}$      &     \,507cm$^{-1}$  \,          \\ \hline
     $\omega_0$\,&\,      42000s$^{-1}$     &     \,94000s$^{-1}$ \,          \\ \hline
     $\alpha_0$\,&\,      1300s$^{-1}$      &     \,4500s$^{-1}$ \,           \\ \hline
     $\delta\omega$\,&\,  225s$^{-1}$       &     \,140s$^{-1}$\,          \\ \hline
     $\delta\alpha$\,&\,  254s$^{-1}$       &     \,145s$^{-1}$\,           \\ \hline
     $\delta\Phi$\,&\,    0.12              &     \,0.023 \,             \\ \hline
\end{tabular}
\end{center}
\caption{$\omega_0$ and $\alpha_0$ are the values of the
phenomenological frequency and damping parameters fitted on a
clean water surface with
 Eq.~\ref{dampedcosine}, for two different
experiments that represent the range of conditions encountered.
$\delta\omega$, $\delta\alpha$ and $\delta\Phi$ are the spread of
the fitted values over 10 independent correlation function
measurements at identical conditions, and they are the best
indicator of  the intrinsic experimental error. The values
$\delta\omega$, $\delta\alpha$ and $\delta\Phi$ are used, as
described in the text,  to estimate the valid fitting regions
which are drawn in Figure~\ref{fitting region}. \label{table
errors}}
\end{table}

%
%
%\\ FIGURE -good fitting region\\
\begin{figure}[tp]
    \centering
        \epsfig{file=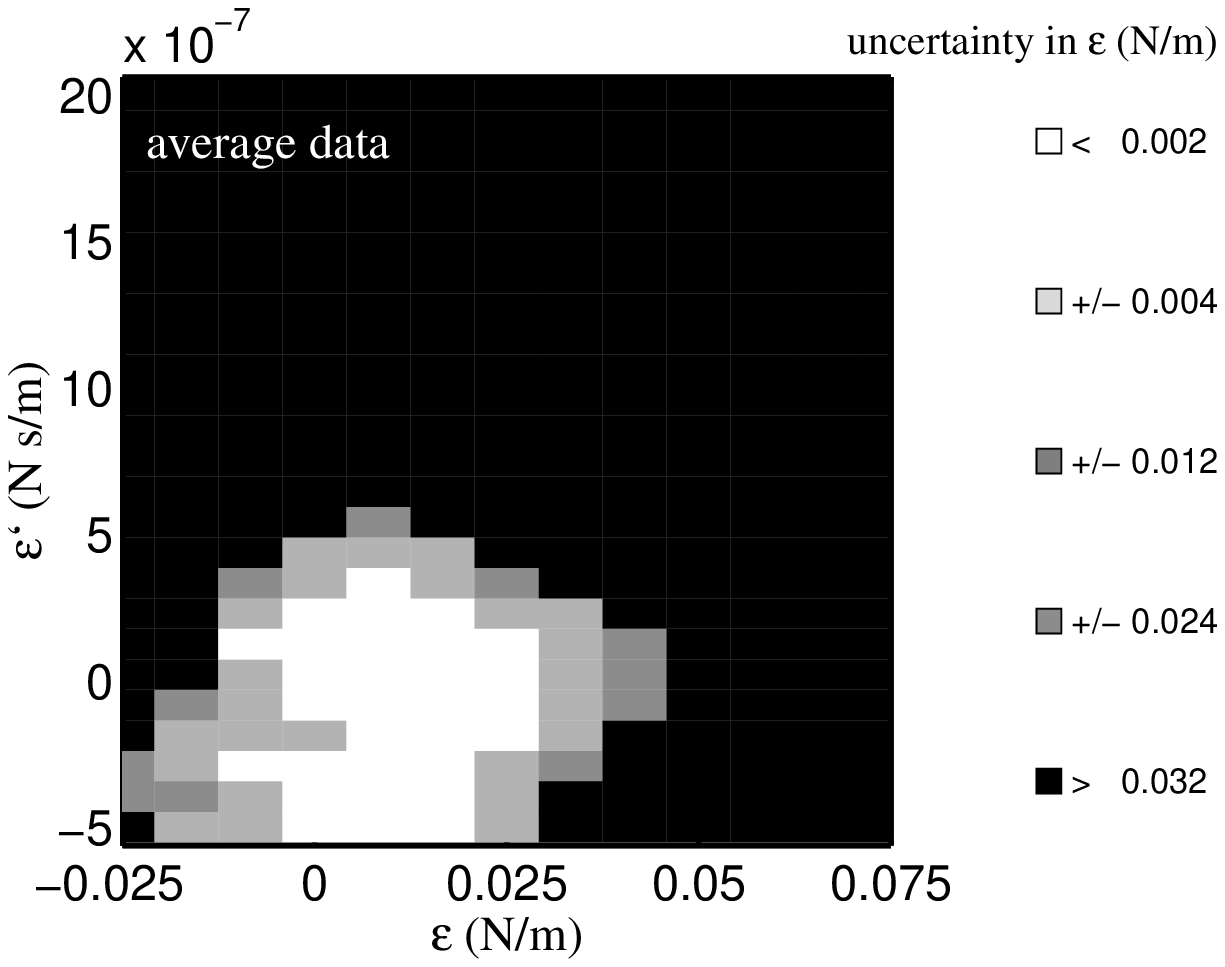,height=5.3cm}
        \epsfig{file=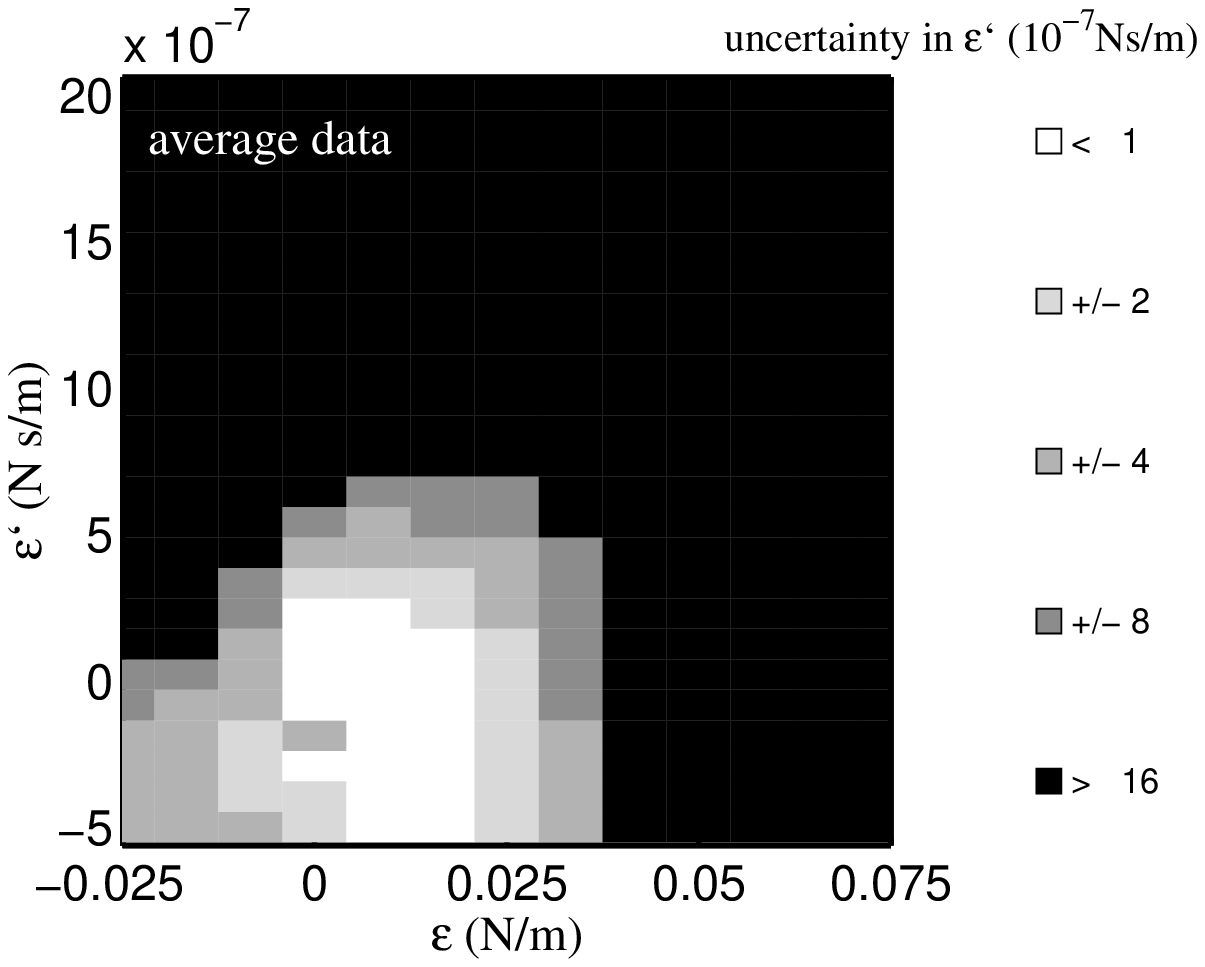,height=5.3cm}\\
          \epsfig{file=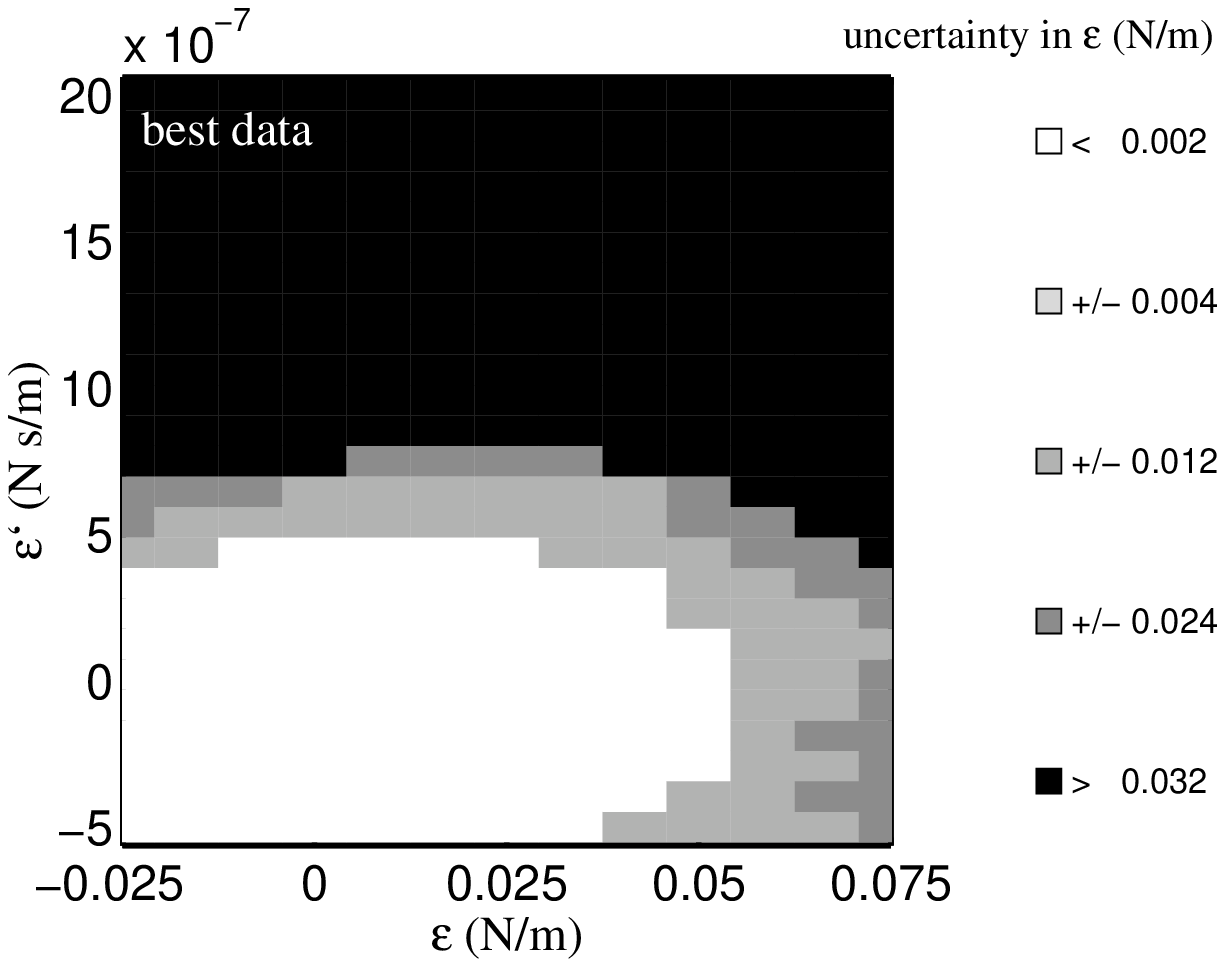,height=5.3cm}
          \epsfig{file=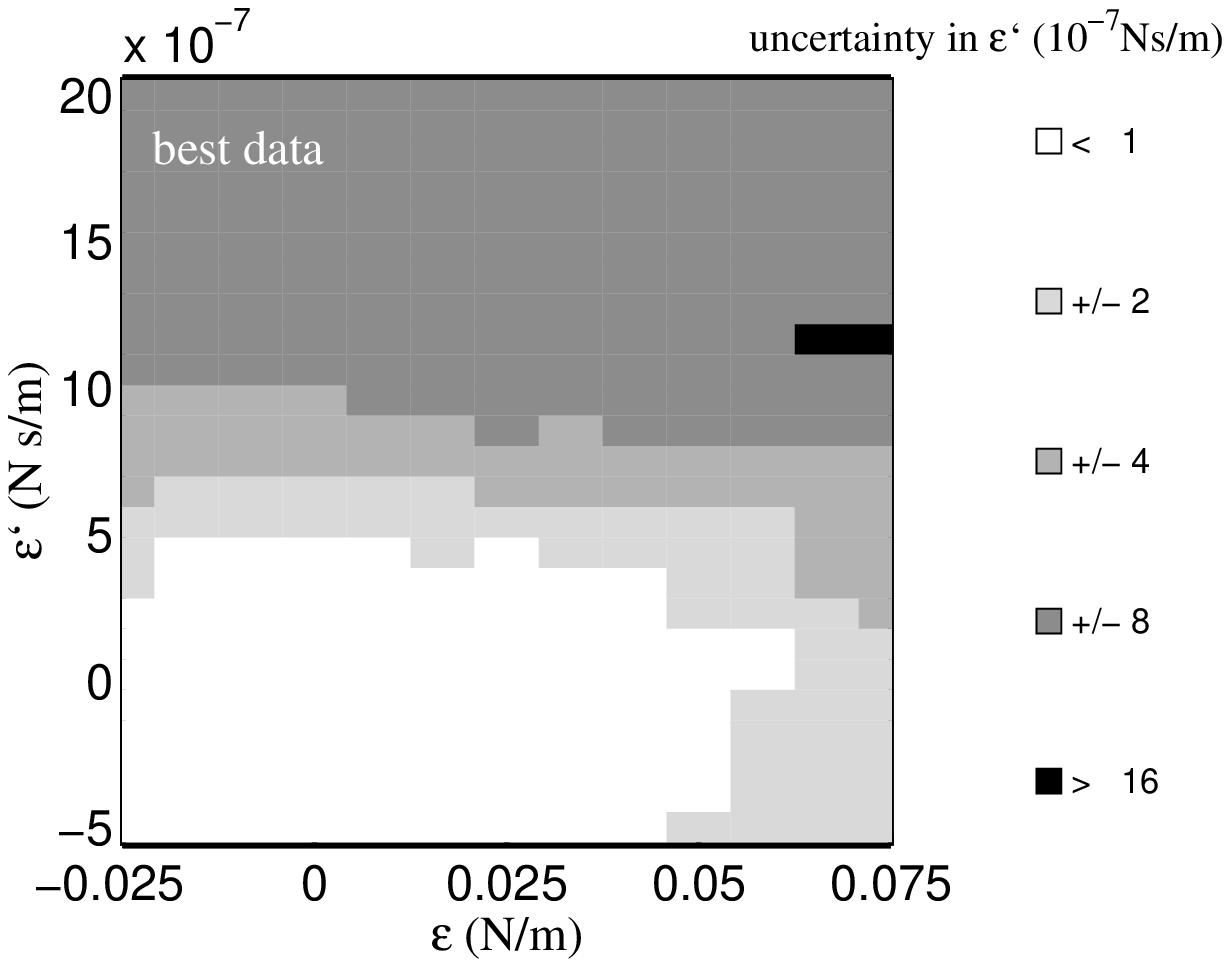,height=5.3cm}
     \caption{Figures showing in white  the region in the
     $(\varepsilon,\varepsilon')$ plane where the SQELS data is expected
     to give reliable fits for $\varepsilon$ (left panels) and for $\varepsilon'$ (right
     panels). As marked on the figures, the top panels have been calculated
     under
      average experimental conditions, at $q=306$cm$^{-1}$ and poor signal to noise ratio,  corresponding to the data
      shown in Figure~\ref{sqels setup figure}(b); The bottom
     panels are under the best conditions achieved with the instrument described here,
     and  correspond to the data shown and discussed in
     ref.~\cite{cicuta01} at $q=507$cm$^{-1}$.
      The legends indicate quantitatively the extent of the $(\varepsilon,\varepsilon')$
      region that will  give
     experimentally      indistinguishable correlation functions.
      The data from which these results are calculated is
     summarized in Table~\ref{table errors}.
\label{fitting region}}
\end{figure}

\subsection{Estimating the valid fitting region}\label{Estimating the valid fitting region} It is generally
acknowledged in the SQELS literature on monolayers that data
analysis fitting the power spectrum with Eq.~\ref{dispeq}
and~\ref{powersp} (or the correlation function with
Eq.~\ref{powersptime}) becomes more difficult as the parameters
$\varepsilon$ and $\varepsilon'$ become large
\cite{yu89b,lan92,yu00}. This is because the power spectrum of
scattered light tends to become independent of  the dilational
moduli in the limit of large $\varepsilon$ or $\varepsilon'$ where
the in-plane and out-of-plane modes will have a very different
characteristic frequency. It is important to establish where is
the boundary of the surface parameter region that can be explored.
Here a method is presented to estimate the range of $\varepsilon$
and $\varepsilon'$ where  the SQELS time correlation data can be
expected to be reliably fitted.

The method of estimating the reliable region of surface parameters
is based on using the stable and reliable fit with the
phenomenological Eq.~\ref{dampedcosine} to give an estimate of the
experimental error that is intrinsic to surface light scattering
data. It can be summarized  in the following steps:
\begin{enumerate}
    \item  Generating model data with Eq.~\ref{dispeq},
    Eq.~\ref{powersp} and Eq.~\ref{powersptime}, with fixed $\gamma$ and over a wide range of
    physical parameters:
    $-100<\varepsilon ($mN/m$)<100$ and $-5< \varepsilon'(10^{-7}$Ns/m$)<20$;
    \item  Fitting the modelled data with Eq.~\ref{dampedcosine},
    thus    establishing three maps: $(\varepsilon,\varepsilon')\rightarrow \omega
    $; $(\varepsilon,\varepsilon')\rightarrow \alpha $; $(\varepsilon,\varepsilon')\rightarrow \phi
    $;
    \item Finding, for each fitted value  $(\omega_{fit}, \alpha_{fit}, \phi_{fit})$, the
    values of $(\varepsilon,\varepsilon')$ that map into a region
     $(\omega_{fit}\pm \delta\omega, \,\alpha_{fit}\pm \delta\alpha, \,\phi_{fit}\pm
     \delta\phi)$, where these error bands are estimates of the experimental error intrinsic to
     the SQELS signal, obtained by measuring the spread in
     the results of  fits with Eq.~\ref{dampedcosine} on 10 separate  sets of real data;
    \item Recording the spread in each of the $\varepsilon$ and
    $\varepsilon'$ sets that map into $(\omega_{fit}\pm \delta\omega, \,\alpha_{fit}\pm \delta\alpha, \,\phi_{fit}\pm
     \delta\phi)$;
    \item Determining the region in the parameter space
    $(\varepsilon,\varepsilon')$ that generates data which will be distinguishable to SQELS.
\end{enumerate}

The results of such an investigation are presented  in
Figure~\ref{fitting region}.  For a realistic range of
$(\varepsilon,\varepsilon')$, Figure~\ref{fitting region} shows
the spread in the fitted values that can be expected when fitting
SQELS data. In practice such figures serve to identify the range
of surface parameters where the technique works well. In all of
the panels, the surface tension $\gamma=65$mN/m and subphase
conditions correspond to water at T$=20^\circ$C.
  Very similar results (not shown) are obtained
for $\gamma=75$mN/m and $\gamma=55$mN/m.  As explained above,  to
generate Figure~\ref{fitting region} the  measured error in the
experimental data is used, and this depends  on various
experimental conditions which determine the signal to noise ratio.
The data used to generate the two conditions represented in
Figure~\ref{fitting region} is summarized in Table~\ref{table
errors}. The largest experimental error bands correspond to the
poorest
 quality data that can still be usefully analyzed, such as shown in Figure~\ref{sqels
setup figure}(b).  The `best' quality error bands correspond to
the data shown in Ref.~\cite{cicuta01}. In general it is clear
that the region of good coupling between the transverse an
in-plane  modes is quite sharply delimited. In particular, it can
be seen that reliable fits for the dilational modulus
$\varepsilon$ can be expected in the worst conditions only for
$\varepsilon<25$mN/m and $\varepsilon'<4\times 10^{-7}$Ns/m. In
the best conditions the region is larger and extends to
$\varepsilon<55$mN/m and $\varepsilon'<5\times 10^{-7}$Ns/m.
Reliable fits for the dilational viscosity $\varepsilon'$ can be
expected in the worst conditions for a quite limited region
$\varepsilon<20$mN/m and $\varepsilon'<3\times 10^{-7}$Ns/m, but
in the best conditions the area increases considerably to
$\varepsilon<55$mN/m and $\varepsilon'<10 \times10^{-7}$Ns/m with
a reasonable confidence
 in the results.

\section{A model polymeric monolayer}\label{A model polymeric monolayer}
Poly(vinyl acetate) (PVAc) is well known as a polymer that forms
surface monolayers which are well described by 'good-solvent'
conditions within the two-dimensional semi-dilute  statistical
model for polymers~\cite{rondelez80}. As commented by Jones and
Richards~\cite{richards99}, PVAc monolayers have been measured
with SQELS by various groups, following initial work by
Langevin~\cite{langevin81}. This makes it an ideal monolayer to
study for comparing experimental and data analysis methods with
the literature. We do not consider studies where data was fitted
attempting to find an imaginary component for the surface tension,
an approach which is now recognized as incorrect~\cite{buzza02}.
The studies that we consider for comparison are from Yu's
group~\cite{yu89b,yu89,yu00} and Monroy's
group~\cite{monroy98,monroy99,monroy00b}. In the experiments of
Yu~{\it et al.} data was obtained over a narrow range of
wavevectors $323<q($cm$^{-1})<445$, while Monroy~{\it et al.}
acquired data for $115<q($cm$^{-1})<804$. Both groups studied the
effect of concentration and temperature on the dilational modulus
and viscosity. While there is a qualitative agreement of all
experiments finding that the SQELS and equilibrium dilational
moduli are the same in the semi-dilute regime, the data of
dilational viscosity differs rather widely. Furthermore,
Monroy~{\it et al.} find a strong dependence on
temperature~\cite{monroy00b}, while Yu~{\it et al.} had found no
such effect~\cite{yu89}. Both groups perform data analysis
following the `phenomenological fitting' approach discussed above.
There seems to be inconclusive data from the different experiments
particularly at high concentration. We are not concerned here with
the possible experimental details leading to such different
behavior in similar systems. Instead, we present new  data on PVAc
that is analyzed by `direct fitting' as discussed in the previous
sections, and we extend the comments on the  valid region for
fitting to discuss the literature data.

\subsection{SQELS on PVAc monolayers}
Experiments are performed under conditions very similar to
ref.~\cite{monroy98}. Solutions of PVAc (Acros 183255000,
M$_w$=170000, used as received) 0.12mg/ml in tetrahydrofurane, are
spread on a pure water subphase, at different temperatures. The
monolayer forms spontaneously and is compressed in a Langmuir
trough. SQELS data is acquired at a given set of concentrations.
The concentration range which is explored is within the
semi-dilute regime, that is for concentrations below the maximum
in the equilibrium dilational modulus~\cite{cicuta01}.

Figure~\ref{figuresqelstam} shows the parameters obtained by a
`phenomenological fit'. They are typical of polymer monolayers
described by the dispersion relation Eq.~\ref{dispeq}. In
particular, the decrease of the frequency $\omega$ with
concentration is due to the reduction of surface tension, and the
peak in the damping coefficient $\alpha$ has been shown to occur
at the resonance condition between in-plane and out-of-plane
modes~\cite{richards99}. Traditionally, it is from these fitted
parameters that the values of dilational modulus and viscosity are
extracted by further analysis. As described above, we follow the
alternative `direct fitting' method. It is extremely important to
accurately measure the subphase temperature and use the
corresponding values of  density and viscosity.
Figure~\ref{figuresqelsbcas2}(b) and~(c) show the dependence of
$\varepsilon$ and $\varepsilon'$ on the concentration. Also shown
on the figures is the limit beyond which the time correlation data
does not depend on $\varepsilon$ and $\varepsilon'$, as calculated
in Figure~\ref{fitting region} for `average data' conditions. It
is clear that the discrepancy above the concentration
$\Gamma_{surf} \simeq 0.6 \times 10^{-3}$g/m$^2$ between the
equilibrium dilational modulus and $\varepsilon$ obtained from
SQELS is most likely due to the parameters being outside the
fitting region. A similar situation occurs for the protein films
studied in~\cite{cicuta01}, at higher values of the moduli because
the raw data quality was better, see Figure~\ref{fitting region}
under `best data' conditions. It is now clear that the large
scattering in results on PVAc reported by Monroy {\it et al.} and
Yu {\it et al.} at high concentration is probably due to the
surface moduli being outside the reliable fitting region.

In the low concentration region, where the PVAc data presented
here can be fitted reliably, we find that $\varepsilon$ is equal
to the equilibrium value and that there is not any significant
effect of temperature. We can compare the values of $\varepsilon'$
to those in the literature and find good agreement with
ref.~\cite{monroy98}, while ref.~\cite{yu89b} reports values that
are lower by almost an order of magnitude.

\section{Conclusions}
Surface light scattering is a non-invasive and sensitive technique
that can be used to study the dilational rheology of polymer
monolayers on a liquid subphase. The experimental setup and
theoretical framework have been described, and the method of data
analysis has been discussed in depth. It has been shown that
reliable fits can be obtained for a limited range of surface
parameters. In particular it has been emphasized that for large
values of dilational modulus, such as will be attained by typical
polymer monolayers at a high monomer concentration or by fatty
acid layers in the condensed phases, it becomes impossible to
determine the magnitude of the modulus from the light scattering
data. Experimental data on PVAc monolayers has been presented here
to give an example of the methods discussed in the paper and to
bring attention to the uncontrolled approximations that lie behind
the widely used `phenomenological fitting' approach.

%\begin{thebibliography}{00}

% \bibitem{label}
% Text of bibliographic item

% notes:
% \bibitem{label} \note

% subbibitems:
% \begin{subbibitems}{label}
% \bibitem{label1}
% \bibitem{label2}
% If there is a note, it should come last:
% \bibitem{label3} \note
% \end{subbibitems}

%\bibitem{}

%\end{thebibliography}

\bibliographystyle{elsart-num}
      \bibliography{methods}

\begin{thebibliography}{10}
\expandafter\ifx\csname url\endcsname\relax
  \def\url#1{\texttt{#1}}\fi
\expandafter\ifx\csname urlprefix\endcsname\relax\def\urlprefix{URL }\fi

\bibitem{buzza95}
D.~M.~A. Buzza, C.-Y.~D. Lu, M.~E. Cates, Linear shear rheology of
  incompressible foams, J. Phys. II France 5 (1995) 37.

\bibitem{lan92}
D.~Langevin, Light Scattering by Liquid Surfaces and Complementary Techniques,
  Dekker, New York, 1992.

\bibitem{earnshaw96}
J.~C. Earnshaw, Light scattering as a probe of liquid surfaces and interfaces,
  Adv. Coll. Interface Sci. 68 (1996) 1.

\bibitem{huang98}
Q.~R. Huang, C.~H. Wang, Effects of viscoelasticity of bulk polymer solution on
  the surface modes as probed by laser light scattering, J. Chem. Phys. 109
  (1998) 6103.

\bibitem{kubota98}
H.~Nakanishi, S.~Kubota, Absence of surface mode in a visco-elastic material
  with surface tension, Phys. Rev. E 58 (1998) 7678.

\bibitem{levich62}
V.~G. Levich, Physicochemical Hydrodynamics, Prentice-Hall, Inc., Englewood
  Cliffs, 1962.

\bibitem{katyl67}
R.~H. Katyl, U.~Ingard, Line broadening of light scattered from a liquid
  surface, Phys. Rev. Lett. 19 (1967) 64.

\bibitem{webb69}
J.~S. Huang, W.~W. Webb, Viscous damping of thermal excitations on the
  interface of critical fluid mixtures, Phys. Rev. Lett. 23 (1969) 160.

\bibitem{meunier69}
J.~Meunier, Diffusion de la lumi\`{e}re par les ondes de surface sur {CO$_2$}
  pr\`{e}s du point critique mesure de la tension superficielle, J. Phys.
  France 30 (1969) 933.

\bibitem{miller96}
R.~Miller, R.~W{\"{u}}stneck, J.~Kr{\"{a}}gel, G.~Kretzschmar, Dilational and
  shear rheology of adsorption layers at liquid interfaces, Colloids and
  Surfaces A 111 (1996) 75.

\bibitem{dimeglio99}
C.~Barentin, C.~Ybert, J.-M. {di Meglio}, J.-F. Joanny, Surface shear viscosity
  of {Gibbs} and {Langmuir} monolayers, J. Fluid Mech. 397 (1999) 331.

\bibitem{fuller99}
C.~F. Brooks, G.~G. Fuller, C.~W. Curtis, C.~R. Robertson, An interfacial
  stress rheometer to study rheological transitions in monolayers at the
  air-water interface, Langmuir 15 (1999) 2450.

\bibitem{zasadzinski02b}
J.~Ding, H.~E. Warriner, J.~A. Zasadzinski, D.~K. Schwartz, Magnetic needle
  viscometer for {Langmuir} monolayers, Langmuir 18 (2002) 2800.

\bibitem{lucassen69}
E.~H. {Lucassen-Reynders}, J.~Lucassen, Properties of capillary waves, Advances
  Coll. and Interface Sci. 2 (1969) 347.

\bibitem{langevin71}
D.~Langevin, M.~A. Bouchiat, Spectrum of thermal fluctuations of a liquid
  covered by a thin film, Comptes Rendus Acad. Sci. 272B (1971) 1422.

\bibitem{buzza02}
D.~M.~A. Buzza, General theory for capillary waves and surface light
  scattering, Langmuir 18 (2002) 8418.

\bibitem{kramer71}
L.~Kramer, Theory of light scattering from fluctuations of membranes and
  monolayers, J. Chem. Phys. 55 (1971) 2097.

\bibitem{buzza98}
D.~M.~A. Buzza, J.~L. Jones, T.~C.~B. McLeish, R.~W. Richards, Theory of
  surface light scattering from a fluid-fluid inetrface with adsorbed polymeric
  surfactants, J. Chem. Phys. 109 (1998) 5008.

\bibitem{earnshaw87}
J.~C. Earnshaw, R.~C. McGivern, Photon correlation spectroscopy of thermal
  fluctuations of liquid surfaces, J. Phys. D: Appl. Phys. 20 (1987) 82.

\bibitem{neuman81}
S.~H\r{a}rd, R.~D. Neuman, Laser light-scattering measurements of viscoelastic
  mono-molecular films, J. Coll. Interface Sci. 83 (1981) 315.

\bibitem{pecora76}
B.~J. Berne, R.~Pecora, Dynamic Light Scattering, Wiley, New York, 1976.

\bibitem{earnshaw97}
J.~C. Earnshaw, Surface light scattering: A methodological review, Appl. Optics
  36 (1997) 7583.

\bibitem{earnshaw88}
P.~J. Winch, J.~C. Earnshaw, A method for rapid data acquisition in
  photon-counting experiments for time-dependent systems, J. Phys. E -
  Scientific Instruments 21 (1988) 287.

\bibitem{hard76}
S.~H\r{a}rd, Y.~Yamnerius, O.~Nilsson, Laser heterodyne apparatus for
  measurements of liquid surface properties-theory and experiments, J. Appl.
  Phys. 47 (1976) 2433.

\bibitem{langevin74}
D.~Langevin, Light scattering from the free surface of water, J. Chem. Soc.
  Faraday Trans I 70 (1974) 95.

\bibitem{shih84}
L.~B. Shih, Surface fluctuation spectroscopy: A novel technique for
  characterizing liquid interfaces, Rev. Sci. Instrum. 55 (1984) 716.

\bibitem{mann84}
J.~A. Mann, R.~V. Edwards, Surface fluctuation spectroscopy: {Comments} on
  experimental techniques and capillary ripplon theory, Rev. Sci. Instruments
  55 (1984) 727.

\bibitem{swofford88}
R.~B. Dorshow, A.~Hajiloo, R.~L. Swofford, A surface laser-light scattering
  spectrometer with adjustable resolution, J. Appl. Phys. 63 (1988) 1265.

\bibitem{earnshaw90b}
J.~C. Earnshaw, R.~C. McGivern, A.~C. McLaughlin, P.~J. Winch, Light-scattering
  studies of surface viscoelasticity: Direct data analysis, Langmuir 6 (1990)
  649.

\bibitem{takagi91}
K.~Sakai, P.-K. Choi, H.~Tanaka, K.~Takagi, A new light scattering technique
  for a wide-band ripplon spectroscopy at the {MHz} region, Rev. Sci. Instrum.
  62 (1991) 1192.

\bibitem{earnshaw93}
C.~J. Hughes, J.~C. Earnshaw, High frequency capillary waves: A light
  scattering spectrometer, Rev. Sci. Instrum. 64 (1993) 2789.

\bibitem{jorgensen92}
T.~M. Jorgensen, A low-cost surface laser light scattering spectrometer, Meas.
  Sci. Technol. 3 (1992) 588.

\bibitem{sawada97}
Z.~Zhang, I.~Tsuyumoto, S.~Takahashi, T.~Kitamori, T.~Sawada, Monitoring of
  molecular collective behavior at a liquid/liquid interface by a time-resolved
  quasi-elastic laser scattering method, J. Phys. Chem. A 101 (1997) 4163.

\bibitem{meyer97}
P.~Tin, J.~A. Mann, W.~V. Meyer, T.~W. Taylor, Fiber-optics
  surface-light-scattering spectrometer, Appl. Optics 36 (1997) 7601.

\bibitem{meyer97b}
W.~V. Meyer, J.~A. Lock, H.~M. Cheung, T.~W. Taylor, P.~Tin, J.~A. Mann, Hybrid
  reflection-transmission surface light-scattering instrument with reduced
  sensitivity to surface sloshing, Appl. Optics 36 (1997) 7605.

\bibitem{trojanek01}
A.~Troj{\'{a}}nek, P.~Krtil, Z.~Samec, Quasi-elastic laser light scattering
  from thermally excited capillary waves on polarised liquid|liquid interfaces.
  {Part} 1: {Effects} of adsorption of hexadecyltrimethylammonium chloride at
  the water|1,2-dichloroethane interface, J. Electroanalytical Chemistry 517
  (2001) 77.

\bibitem{meyer01a}
{J. A. Mann, Jr.}, P.~D. Crouser, W.~V. Meyer, Surface fluctuation spectroscopy
  by surface-light-scattering spectroscopy, Appl. Optics 40 (2001) 4092.

\bibitem{meyer01b}
W.~V. Meyer, G.~H. Wegdam, D.~Fenistein, {J. A. Mann, Jr.}, Advances in
  surface-light-scattering instrumentation and analysis: noninvasive measuring
  of surface tension, viscosity, and other interfacial parameters, Appl. Optics
  40 (2001) 4113.

\bibitem{ferri01}
D.~Magatti, F.~Ferri, Fast multi-tau real-time software correlator for dynamic
  light scattering, Appl. Optics 40 (2001) 4011.

\bibitem{yu02}
D.~G. {Miles Jr.}, Z.~Yang, H.~Yu, Surface light scattering adapted to the
  advanced undergraduate laboratory, J. Chem. Edu. 79 (2002) 1007.

\bibitem{richards99}
R.~A.~L. Jones, R.~W. Richards, Polymers at Surfaces and Interfaces, Cambridge
  Univ. Press, Cambridge (U.K.), 1999.

\bibitem{richards00}
A.~J. Milling, R.~W. Richards, R.~C. Hiorns, R.~G. Jones, Surface viscoelatic
  properties of spread films of a polysilylene-poly(ethylene oxide) multiblock
  copolymer at the air/water interface, Macromolecules 33 (2000) 2651.

\bibitem{monroy98}
F.~Monroy, F.~Ortega, R.~G. Rubio, Dilational rheology of insoluble polymer
  monolayers: {Poly(vinylacetate)}, Phys. Rev. E 58 (1998) 7629.

\bibitem{monroy00b}
F.~Monroy, F.~Ortega, R.~G. Rubio, Thermoelastic behavior of polyvinylacetate
  monolayers at the air-water interface: {Evidences} for liquid-solid phase
  transition, Eur. Phys. J. B 13 (2000) 745.

\bibitem{monroy99}
F.~Monroy, F.~Ortega, R.~G. Rubio, Rheology of a miscible polymer blend at the
  air-water interface. {Quasielastic} surface light scattering study and
  analysis in terms of static and dynamic laws, J. Phys. Chem. B 103 (1999)
  2061.

\bibitem{monroy01}
F.~Monroy, S.~Rivillon, F.~Ortega, R.~G. Rubio, Dilational rheology of
  {Langmuir} polymer monolayers: {Poor-solvent} conditions, J. Chem. Phys. 115
  (2001) 830.

\bibitem{yu89}
K.-H. Yoo, H.~Yu, Temperature dependence of polymer film properties on the
  air-water interface: {Poly(vinyl acetate)} and poly(n-butyl methacrylate),
  Macromolecules 22 (1989) 4019.

\bibitem{yu94}
F.~E. Runge, M.~S. Kent, H.~Yu, Capillary wave investigation of surface films
  of diblock copolymers on an organic subphase:
  {Poly(dimethylsiloxane)}-poly(styrene) films at the air/ethyl benzoate
  interface, Langmuir 10 (1994) 1962.

\bibitem{yu00}
A.~R. Esker, L.~Zhang, B.~B. Sauer, W.~Lee, H.~Yu, Dilational viscoelastic
  behaviors of homopolymer monolayers: {Surface} light scattering analysis,
  Colloids and Surfaces A 171 (2000) 131.

\bibitem{cicuta02}
P.~Cicuta, I.~Hopkinson, Dynamic light scattering from colloidal fractal
  monolayers, Phys. Rev. E 65 (2002) 041404.

\bibitem{cicuta01}
P.~Cicuta, I.~Hopkinson, Studies of a weak polyampholyte at the air-buffer
  interface: The effect of varying ph and ionic strength, J. Chem. Phys. 114
  (2001) 8659.

\bibitem{yu89b}
M.~Kawaguchi, B.~S. Sauer, H.~Yu, Polymeric monolayer dynamics at the air/water
  interface by surface light scattering, Macromolecules 22 (1989) 1735.

\bibitem{rondelez80}
R.~Vilanove, F.~Rondelez, Scaling description of two-dimensional chain
  conformations in polymer monolayers, Phys. Rev. Lett. 45 (1980) 1502.

\bibitem{langevin81}
D.~Langevin, Light-scattering study of monolayer viscoelasticity, J. Coll. and
  Interface Sci. 80 (1981) 412.

\end{thebibliography}

\end{document}